# Operando X-ray characterization of interfacial charge transfer and structural rearrangements

*—————— Accepted Manuscript ——————*

*Reshma R. Rao[1], Iris C. G. van den Bosch[2], Christoph Baeumer[2,3,]* *

1   Department of Materials, Imperial College London, London W12 0BZ, United Kingdom
1   MESA+ Institute for Nanotechnology, University of Twente, Faculty of Science and Technology, 7500 AE Enschede, Netherlands
2   Peter Gruenberg Institute and JARA-FIT, Forschungszentrum Juelich GmbH, 52425 Juelich, Germany

## Abstract

Key technologies in energy conversion and storage, sensing and chemical synthesis rely on a detailed knowledge about charge transfer processes at electrified solid-liquid interfaces. However, these interfaces continuously evolve as a function of applied potentials, ionic concentrations and time. We therefore need to characterize chemical composition, atomic arrangement and electronic structure of both the liquid and the solid side of the interface under operating conditions. In this chapter, we discuss the state-of-the-art X-ray based spectroscopy and diffraction approaches for such "operando" characterization. We highlight recent examples from literature and demonstrate how X-ray absorption spectroscopy, X-ray photoelectron spectroscopy and surface X-ray diffraction can reveal the required interface-sensitive information.

## Keywords

- Operando
- X-ray absorption spectroscopy
- Photoelectron spectroscopy
- Surface X-ray diffraction
- Crystal truncation rod
- Standing wave XPS
- Electrocatalysis
- Electrochemistry
- Termination layers
- Active sites
- Oxygen evolution reaction
- Materials Science
- Surface Science

## Key points/objectives

- True active phases and reaction sites do not pre-exist in as-prepared electrode surfaces, but are formed (and consumed) in a unique surface termination layer that evolves under reaction conditions
- Operando characterization of solid-liquid interfaces is necessary to identify the active surface phases and understand structure-property-function relationships for energy applications
- We describe the fundamentals and application of operando X-ray absorption spectroscopy, including pathways for enhanced interface-sensitivity of a nominally bulk-sensitive technique
- We discuss recent and future developments that enable surface-sensitive characterization of composition and electronic structure using operando X-ray photoelectron spectroscopy
- We present the use of surface X-ray diffraction as a tool to understand surface structure, nature of adsorbates and interfacial electrolyte structure of model electrode-electrolyte interfaces

## Copyright

## 1 Introduction

Fundamental understanding of electronic, atomic and molecular processes at the solid-liquid interface is key for numerous current and emerging technologies such as sensors, chemical synthesis, and energy conversion and storage. These technologies rely on the transfer of electrons, electronic holes and ions across the interface driven by the electrochemical potential differences, i.e. so-called charge-transfer processes, for example during adsorption and desorption of ionic species, and during redox reactions in the solid electrode or in the liquid electrolyte. At the same time, interfaces can present the bottleneck in technologies relying on charge transfer processes. These challenges and opportunities require experimental probes to shed light on the underlying physical and chemical processes to enable rational design of interfaces to optimize electrochemical processes.

The details of the charge-transfer reactions depend on the electronic and molecular structure of the solid-liquid interface. This structure and resulting properties were first described in the 1800s[1–3] and refined in the 20$^{th}$ century.[4] Yet the atomic details are still under discussion and our understanding is continuously improving,[4–6] as also summarized in the previous chapters. The continuing pursuit of deeper understanding is aided by the development of new experimental probes that enable characterization of the molecular-level structure of the solid and the liquid and of dynamic charge-transfer processes.[7] To date, our most complete understanding of the solid-liquid interface has been derived from X-ray characterization (scattering and spectroscopy) of the liquid[6,8] and electronic[5] structure at the interface of noble metal electrodes and aqueous electrolytes. In general, such experiments must be performed "operando" (measurement under working conditions, i.e. while the reaction of interest occurs, with simultaneous measurement of reaction rate[9,10]) because the structure, chemical composition and physico-chemical properties of both the solid and the liquid side of the interface change (reversibly or irreversibly) as a function of external stimuli like applied potentials.

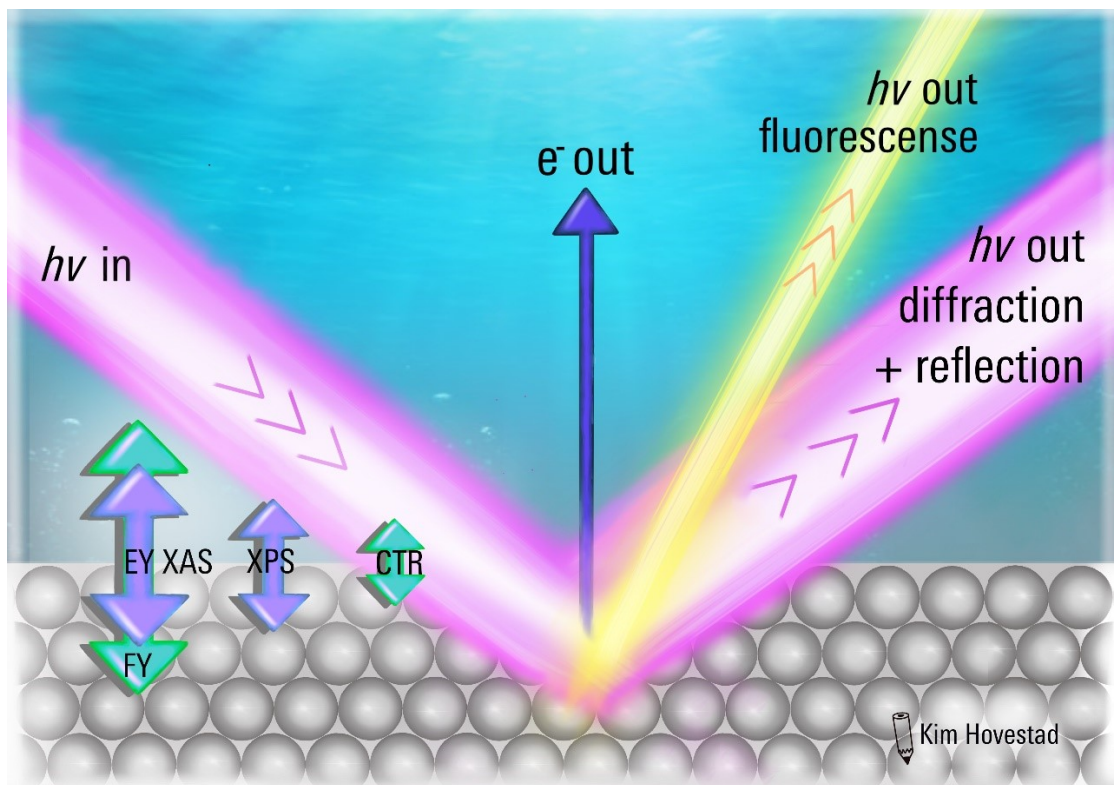

**Figure 1:** Artist's rendering of the solid-liquid interface. X-rays illuminate the sample, leading to scattering, reflection, and emission of fluorescence photons and photoelectrons. The relative information depth is schematically shown by the arrows on the left (arrows not to scale).

The solid-liquid interface under reaction conditions undergoes transformations because of electrochemical processes like ion insertion or surface (redox) reactions such as adsorption, desorption, dissolution and phases changes.[11] These can precede electrocatalytic reactions of interest[12] or occur at almost the same potential.[13] Therefore, both the surface and the bulk of the solid at applied stimulus may have a different composition and structure compared to open-circuit conditions.[14] In addition, an electrostatic double layer forms at the interface when a solid is immersed in a liquid, modifying the potential profile across the interface. Transformations are especially important for more complex solids like transition metal oxides and carbides compared to the relatively well-understood noble metal interfaces because the former exhibit an intricate set of electrochemical phenomena including bulk ion intercalation alongside several coexisting reactions at the interface, which need to be separated by experimental probes under operating conditions.

In general, such transformations are difficult to characterize because several processes coexist in the same materials and under similar conditions, either competing with or assisting one another, and they can be accompanied by a loss of long-range order.[14,15] Since the term "operando characterization" was termed in 2002, the number of experimental studies employing operando characterization has steadily increased, and significant experimental and conceptual progress has been achieved. Yet fundamental experimental challenges remain, especially regarding interface-sensitivity and -selectivity. The performance-enabling or performance-limiting processes are governed by the properties of the interface itself or within nanometer-sized interfacial layers. But experimentalists face small signals from the relevant interface itself, because many experimental probes are either bulk-sensitive (such as most photon-in, photon-out spectroscopies) or not applicable to solid-liquid interface because they require a vacuum (such as regular electron spectroscopies). Therefore, we need interface-sensitive and interface-selective operando probes that collect interpretable signal from the interface of interest without overshadowing by the bulk solid or liquid. Such techniques can be based on enhancing interface-sensitivity for nominally bulk-sensitive techniques, making use of the symmetry-breaking at the interface for operando probes that possess signal only from the interface, or by modifying the experimental design for interface-sensitive techniques that usually cannot accommodate operation with liquid layers.

In this chapter, we discuss X-ray absorption spectroscopy (XAS), X-ray photoelectron spectroscopy (XPS) and surface X-ray diffraction, starting with the most bulk-sensitive method and progressing towards enhanced interface-sensitivity, as schematically depicted in Figure 1 and as summarized in Table 1. Other, undoubtedly also important and promising techniques for the study of various interfacial processes are beyond the scope of this chapter. Several insightful reviews on the topic can be found in references [16–23]. For each technique, we provide an overview about the measurement principles, the necessary cell designs for operando measurements, the experimental measures to achieve or ensure interface-sensitivity and the need for simulation-assistance in interpretation. We summarize intriguing example applications from recent scientific literature regarding aqueous electrocatalytic reactions. A focus is the oxygen evolution reaction (OER), which limits the energy efficiency in various green energy conversion and storage technologies due to low catalytic activities. But the experimental approaches and techniques are also relevant for other technologies relying on charge-transfer across the solid-liquid interfaces, such as batteries and sensors. We will also include our personal perspective on future developments and experimental avenues for improved interface-sensitive X-ray spectroscopic techniques.

Table 1: Selected operando characterization tools

| Technique | Probe/detected species | Sensitive for | Advantages | Disadvantages | Further Reading |
|---|---|---|---|---|---|
| **Sample/experimental design X-ray absorption spectroscopy (XAS) and X-ray photoelectron spectroscopy (XPS)** | | | | | |
| **XAS and XPS with membrane-cells** | Photons in, electrons or photons out | Local structure, atomic concentrations, oxidation states and electrostatic potentials | Avoid limitations from mass transport using a flow cell setup. Can be interface-sensitive | Limitation to selected materials and geometries. Risk of membrane failure | 16,21,24 |
| **Meniscus XPS** | Photons in, electrons out | Atomic concentrations, oxidation states and double layer potential | Solid material of any thickness interfaced with a liquid. Sensitive for band alignments at the interface | Mass and charge transport limitations. Limited information depth or limitations in signal-to-noise ratio. Meniscus instability | 25–29 |
| **Standing wave XPS and XAS** | Photons in, electrons or photons out | Atomic concentrations, oxidation states and double layer potential with extreme depth resolution | Highest depth resolution in X-ray spectroscopy (Ångström-scale) | Complicated samples, long measurement times, complex analysis | 30–33 |
| **Type of X-rays** | | | | | |
| **Hard (5-10 keV)** | Photons in, electrons or photons out | Atomic concentrations, oxidation states, local geometries, and electrostatic potentials | Comparably simple experimental cell, larger information depth | Not interface-sensitive enough. Usually requires synchrotron radiation | 7,34–40 |
| **Soft (50-1500 eV)** | Photons in, electrons or photons out | Atomic concentrations, oxidation states, local geometries, and electrostatic potentials | Very sensitive for oxidation state and local geometry, smaller information depth | Not always interface-sensitive enough. Usually requires synchrotron radiation and complicated experimental setups. Need for UHV. | 7,34–36,41–48 |
| **Surface x-ray diffraction (SXRD)** | | | | | |
| **Surface X-ray diffraction** | Photons in, photons out | Surface structure, adsorbed species, structure of the liquid layer | Very sensitive for the interface structure. Comparably simple cell design (hard X-rays) | Requires extensive modelling and prior knowledge about the surface structure. Limited to highly crystalline and very flat samples. Usually requires synchrotron radiation. | 8,49–55 |

# 2 X-ray absorption spectroscopy

## 2.1 Theory of the technique

XAS can be utilized to probe the unoccupied states of the electronic structure.[34] Upon X-ray illumination of matter, the photon energy is transferred to an electron. The resulting excitation from a core level towards a specific electronic state results in element-specific absorption edges (i.e., sharp increases in absorption coefficient) for photon energies of hundreds or thousands of eV. Fermi's golden rule[56,57] describes the probability of an electron being excited from its ground state with energy $E_i$ by the X-ray energy $h\nu$ to its excited state with energy $E_f = E_i + h\nu$.

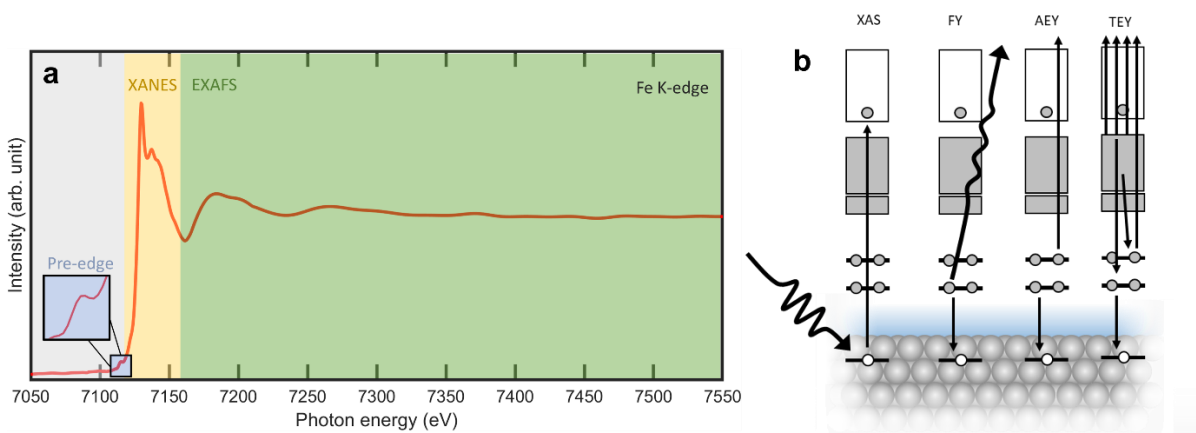

**Figure 2:** (a) Exemplary (smoothed) XAS spectrum of the Fe K-edge, obtained from a 10 nm LaFeO$_3$ thin film. (b) Schematic of the measurement modes of X-ray absorption. Fluorescence yield (FY), Auger electron yield (AEY) and total electron yield (TEY).

Figure 2a shows a typical XAS spectrum of the Fe K-edge, consisting of a pre-edge feature, the X-ray absorption near edge structure (XANES) and the extended X-ray absorption fine structure (EXAFS).[58] The energy region of XANES is defined as the absorption edge and some fine structure up to 50 eV above the edge. Shifts in the absorption edge can be observed when the oxidation state changes, where higher oxidation states result in a shift to higher photon energies due to shielding effects.[58] In addition to the oxidation state, XANES contains information regarding coordination geometry and number, and elemental composition, which can be extracted from peak shapes, positions, intensities, and spin-orbit splitting. A pre-edge feature can be observed in the K-edge of transition metal spectra. Transition metals have a partially occupied $d$ orbital and transitions from $1s$ to $(n-1)d$ states can occur. This transition is formally dipole forbidden, hence the weak intensity in comparison with the edge (transition from the $1s$ to $np$ orbital).[59] The EXAFS region, i.e. the photon energy region after the XANES region, provides information on the distance of the neighboring atoms and crystallographic arrangement. This structural information is obtained from the constructive and destructive interference resulting from scattering of the excited electron with neighboring atoms.

In XAS, the information depth depends critically on the detection mode. Traditionally, XAS is measured in transmission mode, where a thin sample is penetrated by X-rays. Based on the absorption coefficient of the elements an absorption spectrum can be obtained by measuring the attenuated signal after the sample.[60] While measuring in transmission mode, there is a large restriction on the sample thickness. The attenuation length of most materials is around a few micrometers, so the total thickness of the transmission cell is limited to a few tenths of micrometers. So, this is a bulk sensitive measurement technique with hundreds of nanometers to few micrometers information depth.

Indirect methods to measure the absorption spectra are fluorescence yield, Auger electron yield and total electron yield,[57] as shown in the schematic in Figure 2b. The fluorescence yield makes use of the photons that are emitted upon relaxation from the excited state. The emission intensity is measured at each incidence energy. Like transmission mode, this is a bulk sensitive technique (~hundreds of nanometers). The effective penetration/information depth can be decreased using grazing incidence (or grazing exit) geometries. At angles α between the surface tangent and the incoming beam, the X-ray penetration depth decreases with $\sin(\alpha)$. The absolute values of the penetration depth can be calculated based on the material-specific and energy-dependent X-ray absorption coefficients, as tabulated by Henke et al.[61] For example, an information depth of 2 nm was achieved for Pt electrodes and 4 nm for the perovskite oxide $La_{0.6}Sr_{0.4}MnO_3$ for grazing angles of 0.27 and 1° respectively.[41,62]

Surface sensitivity can also be achieved through electron detection because the escape depth of electrons is much smaller (1-10 nm) compared to the fluorescence mode.[35] For total electron yield, all the electrons measured by a current collector are used, resulting in a near-surface sensitive information. Another measurement mode is partial electron yield, where an electron analyzer is used to measure emitted electrons with a specific energy, for example Auger electrons, which can enhance surface sensitivity even further.

## 2.2 Experimental measurement

X-ray absorption spectroscopy experiments are usually performed at synchrotron facilities, and a growing amount of beamlines facilitate operando studies.[7] To study operando solid-liquid interfaces, using hard X-rays is easiest. The energy is typically 5-10 keV, resulting in an attenuation length of up to hundreds of micrometers. High photon energies of thousands of electron volts are used and ultra-high vacuum is not required. Hard X-rays can be used to measure the K-edges of 3$d$ transition metals, from which information about the oxidation state can be obtained based on the edge position.[63] On the other hand, with soft X-rays, with a typical energy of 50-1500 eV, L-edges of these metals can be measured. These are more sensitive to the oxidation state and details of the electronic structure. In addition, light elements like Li, C, N and O only have absorption edges in this regime. Using soft X-rays makes operando measurements more challenging, due to the smaller attenuation lengths of ~1 μm.

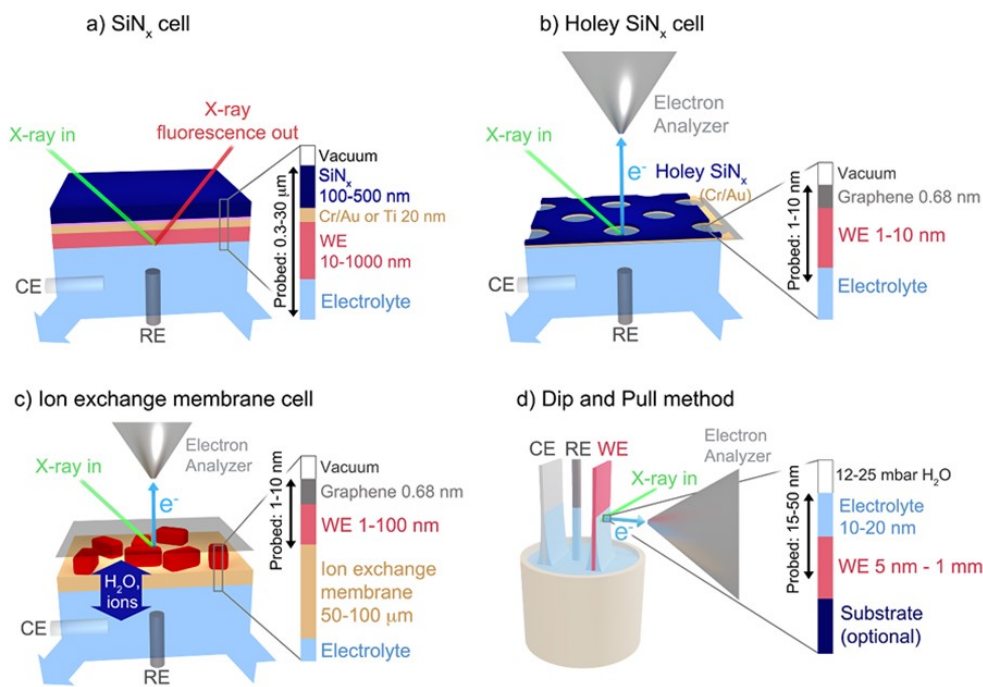

**Figure 3** Cell designs for operando XAS. Reprinted (adapted) with permission from Velasco-Vélez *et al. J. Phys. D. Appl. Phys. 2021, 54 (12), 124003*. (a) A SiN$_x$ flow cell. (b) A holey SiN$_x$ enabling partial electronic yield. (c) An ion exchange membrane. (d) The dip and pull method for XPS measurements. With a counter electrode (CE), a reference electrode (RE) and a working electrode (WE).

Common cell designs are summarized in Figure 3. The SiN$_x$ cell, the holey SiN$_x$ cell and the ion exchange membrane cell are most suitable for X-ray absorption spectroscopy experiments, the dip and pull method will be discussed in section 3.2. The SiN$_x$ cell[64] is a common cell design, where the working electrode is deposited on an X-ray transparent window, often SiN$_x$, with ~100 nm thickness. This cell design was used in some of the pioneering works by the group of Salmeron.[6,7,65] The solid-liquid interface is probed via X-rays entering and exiting through the window and the working electrode. The fluorescence yield mode can be used for this cell design, resulting in bulk information depth in the micrometer range. A transmission geometry with two SiN$_x$ membranes can also be used.[63] The total electron yield can be measured as well through the drain current at the working electrode. The other cell designs are based on the use of thinner, ideally two-dimensional materials, such as graphene, as pioneered by the groups of Kolmakov[66,67] and Salmeron[16,68]. The holey SiN$_x$ cell (Figure 3b) consists of the SiN$_x$ membrane with holes, a graphene layer covers the membrane including the holes. The working electrode (1-10 nm) is deposited on the graphene as thin film or nanoparticles, such that partial electron yield becomes possible, leading to relatively more information from the solid-liquid interface compared to the regular SiN$_x$ cell. Recently, graphene membranes were combined with polymer electrolyte membranes to enhance stability of confined solid-liquid interfaces.[69] For this ion exchange membrane cell (Figure 3c), the working electrode is deposited directly on the polymer electrolyte membrane. A double layer of graphene is deposited on the working electrode to minimize water evaporation into the vacuum chamber. In addition, a thin liquid layer is formed around the working electrode. Similar to the holey $SiN_x$ cell, partial electron yield detection can be used.[64] While tremendous improvement in experimental approaches and valuable operando-insights have been obtained in recent years, all these cell designs for in-situ measurements have some limitations regarding the working electrode material. In general, only nanoparticles and comparably thin films can be used for these cells types, because both incoming and outgoing signal penetrate through the electrode itself.

In addition to the cell designs, the sample design is important for measuring the solid-liquid interface. Let us consider a Gedankenexperiment with $LaNiO_3$ as OER electrocatalyst, measured in fluorescence yield, resulting in bulk-like information depth. It is known that only the top one to two unit cells show chemical and structural changes during operation.[14,70] Only a small contribution of the measured signal is related to this thin surface layer, which depends on the thickness of the catalyst. To overcome this, one can think of designing electrodes with higher surface to bulk volume ratios, for example by using highly porous 3D material or nanoparticles. For now, we put the focus on the latter, as an attractive pathway to increase the surface sensitivity.[27] The thin surface layer on a sphere of 10 nm diameter results in a larger spectral contribution compared to the surface layer of a 10 nm thin film. Operando measurements using nanoparticles have already shown that this pathway creates more interface-sensitive measurements.[71] Still the question remains how to subtract this surface layer data from the total measured spectrum, containing a large percentage of bulk-like information. Liang, Chueh and coworkers have designed and demonstrated an approach to tune the nominally bulk-sensitive techniques for interface-sensitive measurements using epitaxial thin films of various thicknesses.[14] For these experiments, the thin film surface layer has identical intensity for all thicknesses, whilst the bulk contribution increases with increasing film thickness. Even though nanoparticles are relatively more surface sensitive, from thin films the interface sensitivity can be extracted relatively easily, especially for well-defined, epitaxially-grown model systems. However, thickness-dependent properties and varying amounts of defects can give additional challenges in analysis of the data.

## 2.3 Examples of interface-sensitive XAS measurements

The application of operando XAS measurements in electrocatalysts[34,72], solar energy materials,[73] and lithium-ion batteries[35] have been discussed is recent reviews, therefore only selected examples will be discussed briefly.

### A. Change in oxidation state of metal oxide compounds

The interest for operando XAS studies for electrocatalysts promoting the oxygen evolution reaction are increasing rapidly, in order to get a better understanding on the active site during the catalysis. Tracking the oxidation state of transition metal oxides and atomic distances might reveal some details for the still not well understood mechanism of the oxygen evolution reaction. Fe-Ni compounds are an exemplary highly active electrocatalyst. But there is not yet a clear answer to which element is the active site in these compounds and how this evolves during operation. Friebel *et al.*[74] have investigated the (Fe, Ni) oxyhydroxide, the structure model is shown in Figure 4c. From electrochemistry measurements it is shown that pure FeOOH and NiOOH have a lower OER activity compared to the mixed Fe-Ni compound. It is debatable whether the doped Fe-sites are the catalytically active sites. To better understand what happens by Fe doping of NiOOH, XAS spectra were measured on samples with 25% Fe and 75% Ni. The transition metal oxidation state was determined from the XANES region and EXAFS was used to determine the oxygen-metal and metal-metal bond distances under operating conditions. A silicon nitride window covered with a Ti adhesion layer and a Au layer was used. The Ni-Fe catalyst was deposited on top of this. XAS was measured in high energy resolution fluorescence detection mode, which reduces the core-hole lifetime broadening and enhances pre-edge features.

Both the Fe and Ni K-edge spectra exhibit potential induced changes in the XAS spectra. For the Ni K-edge a clear shift in edge and pre-edge can be observed, as shown in Figure 4d. The changes in the Ni spectra can be related to the α-Ni(OH)$_2$ phase at low potentials and γ-NiOOH at high potentials, indicating an oxidation of Ni from 2+ to a 3+ or even mixed 3+/4+. The potential related changes in the Fe spectra cannot be clearly related to a different Fe oxidation state, otherwise the shift should have been more noticeable.

However, analysis of the EXAFS of both the Ni and Fe K-edge reveals a reduced metal-oxygen (M-O) and metal-metal (M-M) bond length at high potentials, as shown in Figure 4b. A third peak at twice the distance of the M-M bond indicates that the Fe atoms are substituted on the Ni-sites and not intercalated between the NiOOH sheets. Often a decrease in bond length is associated with an increase in oxidation state. For the Ni this is clearly related to the shift of the edge towards higher photon energies. The change in bond distance for the Fe could also be related to an increase in oxidation state. However, this was not confirmed by a shift in the edge, which is practically absent. There were small changes observed in the pre-edge region, which could be indicative of a small amount of Fe$^{4+}$. From the Ni/Fe ratio dependency the authors found that the Fe-O bond decreases only in the case of low Fe concentrations (25% and smaller). When the Fe concentration is larger the Fe is no longer substituted in the NiOOH structure but starts to form the catalytically inactive γ-FeOOH phase.

With DFT calculations Friebel *et al.* argue that the active sites for the oxygen evolution reaction are the Fe-sites and not the Ni-sites. Bates *et al.*[75] report similar trends regarding the oxidation state of Ni and Fe and also suggest that Fe is the active site. In addition Bates *et al.* [75] performed XAS studies on Fe-Ni-Co compounds, where they found that Fe stabilizes the Ni$^{2+}$ oxidation state and Co the Ni$^{3+}$ oxidation state under applied potentials. The addition of Co in the lattice promotes the formation of the conductive NiOOH phase by decreasing the bond lengths Ni-O and Fe-O, which in turn lowers the overpotential for the oxygen evolution reaction on the Fe sites. Ismail *et al.*[76] found similar XAS results as Friebel *et al.* and Bates *et al.* for photoelectrochemical water splitting using hematite/Ni(Fe)OOH. Other studies found divergent results. For example, Wang *et al.* showed that in addition to the oxidation of Ni, an increase in oxidation state of Fe from 3+ to 4+ can also be observed.[77]

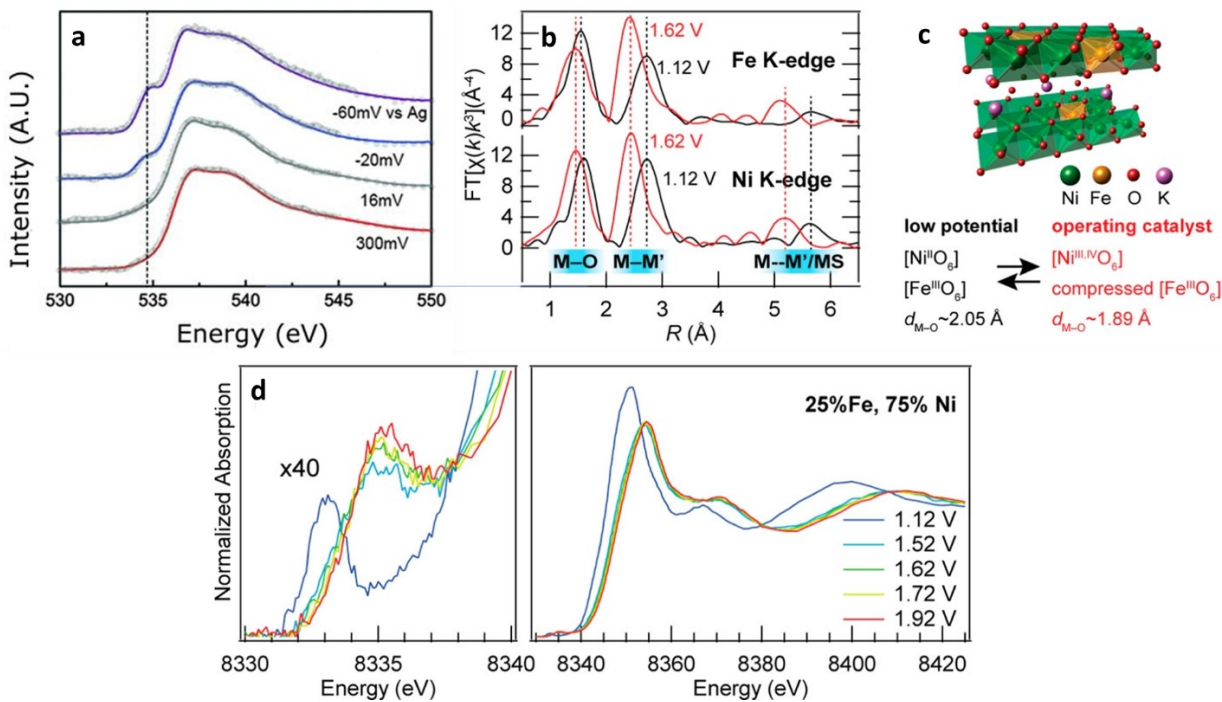

**Figure 4** (a) O K-edge of water molecules at a gold surface in TEY mode, adapted from **V**elasco-Velez *et al.* The Structure of Interfacial Water on Gold Electrodes Studied by X-Ray Absorption Spectroscopy. *Science* **2014**, *346* (6211), 831–834. Reprinted with permission from AAAS. (b) The Fe and Ni K-edge EXAFS region of a (Fe,Ni) oxyhydroxide electrocatalyst. Both show reduced bond lengths at higher applied potentials. (c) The model of the Fe doped NiOOH structure. (d) Ni K-edge spectra, showing a shift in the edge and pre-edge indicating a changed oxidation state. b-d adapted with permission from from Friebel *et al.,* Identification of Highly Active Fe Sites in (Ni,Fe)OOH for Electrocatalytic Water Splitting. *J Am Chem Soc* 2015, *137* (3), 1305–1313. Copyright (2015) American Chemical Society.

### B. Configuration of water at interfaces

In situ measurements of the solid-liquid interface can be either focused on the liquid, solid or both. Here, we discuss the seminal study of water molecules absorbed on a gold surface by Velasco-Velez *et al.* [6] Water molecules form a bias dependent electrical double layer at the interface, related to the strong dipole of water. The bias-dependent changes were studied by operando measurements of the O K-edge, using surface sensitive (TEY) and bulk sensitive (TFY) measurement modes. In this study an electrochemical flow cell was used, with a 20 nm gold layer deposited on a 100 nm thick $Si_3N_4$ membrane window as working electrode (similar to the schematic shown in Figure 3a).

First, the signals of the two detection modes were compared. The information depth of TEY is up to 3 layers of water, based on the attenuation length of the electrons excited from the oxygen K-edge. From ab initio molecular dynamics (AIMD) simulations the possible configurations of water molecules on the surface were simulated as an input for XAS spectra simulations. From simulations of the water matrix closer to the interface, more broken H-bonds are expected compared to the bulk. The measurements showed a reduced pre-peak feature in the oxygen K-edge for TEY measurements related to those broken H-bonds, in contrast to TFY measurements where a distinct pre-edge feature was observed.

Secondly, the expected bias dependency of the water molecule orientation was investigated. Indeed, the pre-edge peak became prominent at negative biases as shown in Figure 4a, whilst at positive biases it was absent. Positive biases increase the amount of dangling bonds oriented towards the gold surface, resulting in a suppression of the pre-edge peak due to coupling with the gold surface. So, this work demonstrated that XAS can reveal structural details of the electrolyte at the solid-liquid interface, and that the bulk information can be compared to the interface sensitive information using different detection modes.

Van Spronsen *et al.* have investigated the solid-liquid interface of $TiO_2$ and water as well, where they found a more ice-like water configuration.[78] This was possible through comparison of the O K-edge of the $TiO_2$-water interface (after subtraction of the O K-edge of $TiO_2$) to reference spectra taken on ice. For this purpose, the authors employed a relatively new measurement mode for membrane cell designs, so-called total ion yield, developed by Schön *et al.*[79] In this approach, the current at the counter electrode is measured. Because charge neutrality is maintained, so electrons going away from the working electrode are measurable at the counter electrode due to ionic transport through the electrolyte, only resulting in a negative signal for the total ion yield compared to the total electron yield.

## 2.4 Need for simulations

The field of operando spectroscopies is growing to uncover the mechanisms in electrochemical reactions. To get a better understanding of the complex results measured by operando XAS, models and simulations are required because the measured spectra are difficult to interpret, for example due to lack of reliable experimental references. In addition, under operando conditions the system is often far from equilibrium, resulting in non-Gaussian distributions of the bond lengths or the presence of slightly different structures. Timoshenko and Roldan Cuenya describe more challenges in ref [34]. Streibel *et al.*[80] propose a combination of operando measurement and calculations to build models of these complex systems. First principal calculations often form the basis for such calculations. The computational hydrogen electrode (CHE) method, first demonstrated by Nørskov *et al.*[81] in 2004, calculates the stability of intermediates and combines this with density functional calculations to obtain a free-energy landscape of catalysts for electrochemical reactions. Density functional theory (DFT) in combination with the Bethe-Salpeter equation (BSE) can be used to calculate theoretical XAS spectra. However, differences between experimental and simulation spectra are observed for transition metals. Another possibility is to use DFT in combination with dynamic mean field theory, which does not show these differences between experiment and simulation regarding the transition metals.

For simulation of the XAS, there are various programs available, for example FEFF, CTM4XAS and Quanty. The latter [82] can be used to calculate x-ray spectroscopy spectrums with multiplet ligand field theory. In addition, it can also do calculations based on non-local interactions. CTM4XAS[83] is a semi-empirical model based on three theoretical components, atomic multiplet theory, crystal field theory and charge transfer theory. Lastly, FEFF[84,85] is a program based on multiple scattering calculations, which can be used for both EXAFS and XANES. It is strongly recommended to do simulations with one of these programs, to help with gaining a better understanding of the measured spectra.

# 3 X-ray photoelectron spectroscopy

## 3.1 Theory of the technique

XPS is sensitive for the electronic structure and chemical composition.[39] In this section, we cover the essentials necessary for the discussion of interface-sensitivity and operando characterization with XPS. A more complete description of the physical process of photoemission and the general application of XPS can be found in seminal review articles[42–44] and in the recent XPS tutorial series collected by the American Vacuum Society with important instructional works by Baer et al.,[45] Powell,[46] Chambers et al.,[47] Tougard[48] and others.

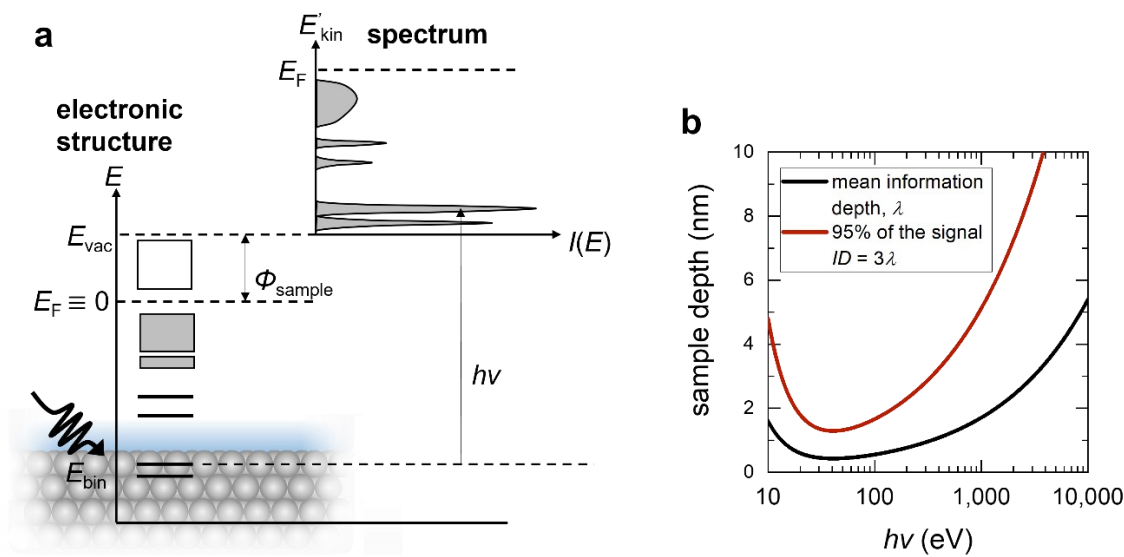

**Figure 5:** (a) Principle of the photoemission process in XPS. Electrons excited by the photons with sufficient energy to leave the sample form a photoelectron spectrum consisting of sharp emission lines from the core levels and a broad valence band distribution (shaded area). (b) Universal curve of the photoelectron inelastic mean free path, λ, as a function of electron kinetic energy. Red line shows the sample depth from which 95% of the total signal originates.

Similar to XAS, XPS relies on the absorption of photons by electrons. The photoexcitation process results in directly measurable electron count rates if the photon energy is higher than the energy difference between a given core level and the vacuum level; a photoelectron leaves the sample and can be detected (Figure 5a). The resulting electron kinetic energy $E'_{kin}$ is

$$E'_{kin} = h\nu - E_{bin} - \Phi_{sample} \qquad 1$$

with the binding energy of the initial core level $E_{bin}$ (i.e. the energy difference between the core level and the Fermi level), the photon energy $h\nu$ and the sample work function $\Phi_{sample}$. If the sample shares a common ground with the XPS analyzer,[42,44] one can use $E'_{kin} + \Phi_{sample} = E_{kin} + \Phi_{analyzer}$, to rewrite equation 1 to:

$$E_{kin} = h\nu - E_{bin} - \Phi_{analyzer} \qquad 2$$

with $E_{kin}$, the electron kinetic energy as measured by the analyzer, and the analyzer work function $\Phi_{analyzer}$. This is convenient because $\Phi_{sample}$ is typically unknown but $\Phi_{analyzer}$ can be easily calibrated. The resulting spectrum is thus converted onto the binding energy scale from the measured kinetic energy of the photoelectron. The peak position on the binding energy scale and the peak shape contain element-specific information about the valence state and electronic structure. Different binding energies are found for a given core level for different oxidation states because of different screening of the nucleus. This so-called chemical shift can be used to identify the valence state. The relative atomic concentrations can be extracted from the integrated intensities, after normalization with relative sensitivity factors that account for differences in the cross sections for the photoelectric effect for different elements, orbitals and instrument geometries.

As electrons are the detected species, XPS is generally a surface-sensitive technique. Photoelectrons are excited within a depth of several hundred nanometers, but as they propagate to the surface, the electrons scatter elastically and inelastically. Inelastically scattered electrons lose energy on their way to the sample surface and thus do not contribute to the characteristic peak intensity but rather contribute to the background of the spectrum, if they leave the sample at all. Neglecting elastic scattering effects, the probability of leaving the sample with the characteristic kinetic energy can be expressed as

$$I(t) = I_0 \, exp \frac{-t}{\lambda_i \cos \theta} \qquad 3$$

with the attenuated intensity $I(t)$ of photoelectrons generated at a depth $t$, the photoelectron intensity without attenuation $I_0$, the photoemission angle $\theta$ (measured between the surface normal and the detector), and the inelastic mean free path $\lambda_i$. In this definition, $I_0$ contains the absorption cross section and the analyzer transmission function for a given energy level. A more complete description was provided by Powell.[46] The total intensity of an atomic species with volume density $\rho$ for a given photoemission angle can then be expressed as

$$I(\theta) = I_0 \int_0^\infty \rho(t) \, exp \frac{-t}{\lambda_i \cos \theta} \, dt. \qquad 4$$

$\lambda_i$ is defined as the "average distance that an electron with a given energy travels between successive inelastic collisions."[46] This depends on the electron kinetic energy and therefore on the chosen X-ray energy and the binding energy of the core level, see equation 1. The absolute $\lambda_i$ values are tabulated in the NIST databases (National Institute of Standards and Technology) and can be predicted using the so-called TPP-2M formalism.[46] For most materials, $\lambda_i$ roughly follows a "universal curve" (Figure 5b), which has a minimum of ~0.3 to 0.4 nm at around 50 eV, and which increases towards higher and lower energies due to decreasing scattering rates with maximum values around 5 nm within the experimentally accessible photon energies. The underlying mechanisms and dependencies on material properties and experimental geometries are still subject of intense research.[46,86–88]

As a result of the small $\lambda_i$ values, XPS is a surface sensitive technique. In practice, it is useful to define quantitative values that describe the surface sensitivity of the XPS experiment with a given material and instrument configuration. Common descriptions are the mean escape depth $\Delta = \lambda_i \cos\theta$ (the average depth normal to the surface from which photoelectrons originate in the given experiment) or the "information depth" $ID = 3\lambda_i$ (the sample depth from which 95 % of the XPS signal originate), again neglecting elastic scattering (Figure 5b). Typical $\Delta$ values range from 0.3 nm to 2 nm for soft X-ray excitation at normal photoemission ($\theta = 0$) and 2 to 6 nm for hard X-ray excitation. In summary, the photoemission angle and excitation energy are decisive for the information depth in XPS, and varying one of the two variables allows for non-destructive depth-profiling within a nm-thin near-surface region.

## 3.2 Experimental measurement

The short inelastic mean free path of photoelectrons implies that typical XPS instrumentation generally requires UHV conditions. Otherwise, the electrons would not reach the detector due to inelastic scattering. But recently, so called near-ambient pressure (NAP- or AP-)XPS tools were developed and commercialized. The approach originally explored by Siegbahn et al. in the 1970s and 1980s[89,90] was perfected at the Advanced Light Source (ALS) in Berkeley and at BESSY in Berlin at the beginning of the 2000s. Now, tens of mbar operating pressures are routinely achievable.[91–94] Key to this development was the introduction of differential pumping stages that progressively reduce the pressure, separating a high-pressure volume near the sample from the required vacuum in the electron analyzer. Together with a short distance between the sample surface and the analyzer condenser lens, this minimizes the scattering probability for the electrons traveling through the high-pressure region and prevents arcing in the electron analyzer at elevated pressure. With further development and even commercial availability of laboratory-based setups,[95,96] NAP-XPS has become a major trend in surface science and the investigation of solid-gas interfaces.[17] Extensive summaries of NAP-XPS for the investigation of the chemical and electronic structure at solid-gas interfaces are provided in refs. [17,94,97–99].

The NAP-XPS approach was also key for the development of XPS characterization tools for the solid-liquid interface, a new research direction that has become very popular in the past few years with rapid installation of dedicated instrument endstations at several synchrotron facilities around the world.[29,100–103] XPS of the solid-liquid interface allows studying electrochemical processes like specific adsorption of ions, charge transfer dynamics, electrical contact potential profiles and compositional changes with time and applied potential. Two main approaches can be distinguished: Meniscus XPS (also called dip and pull XPS) and membrane XPS, both of which will be described below.

### A. Meniscus XPS

The dip and pull approach was pioneered at the Zhi Liu's group at the ALS and relies on the observation that once a hydrophilic solid is partially immersed in an aqueous electrolyte, a stable meniscus may form, as already discussed extensively by Bockris and Cahan in the late 1960s.[104] NAP-XPS chambers allow installation of an open container with liquid electrolyte, where the partial pressure of the solvent in the chamber equals its vapor pressure for the experimental temperature. The ALS team found that immersing and partially extracting the sample from the liquid solution ("dipping" and "pulling"), as shown in Figure 3d and as described in detail in refs. [25–29] results in a stable and photoelectron-transparent meniscus. The electrolyte thickness on a Pt electrode was in the range of 10 nm to 30 nm, which can be penetrated by photoelectrons, especially when using "tender" X-rays of 3-4 keV, as described in the seminal paper by Axnanda, Crumlin et al.[25] Alternative approaches are currently explored, including the "tilted sample"[105] and the "offset droplet" method using a fine capillary.[106,107] These might offer advantages regarding the proximity of the "bulk liquid" but will not be discussed here.

Despite these advantages and achievements, meniscus XPS also faces serious limitations: First, the thin meniscus layer leads to mass transport limitations, because the electrolyte resistance in the meniscus is more than three times higher than in the bulk, even for high electrolyte concentrations.[108] As a result, only low current densities (~below 1.0 mA cm$^{-2}$)[108] can be measured with reasonable IR drop. Even then, the measurement spot might have a different potential than applied, due to possibly non-uniform potential along the length of the electrode.[7] It is therefore necessary to always measure the relative position of electrode and electrolyte core level binding energies, e.g. in the O 1$s$ core level for oxide surfaces measured in an oxygen-containing electrolyte, such as water. The energy difference between the solid (which is fixed to the analyzer potential for sufficiently conductive samples) and the liquid should shift proportionally to the applied potential.[5] Second, the nm-thin meniscus may suffer instabilities due to, among others, influence of gravity, slow loss of electrolyte in a backfilled chamber and higher relative pumping speed in close proximity of the energy analyzer cone. Moreover, unfavorable applied potentials may lead to shrinking of the stable meniscus,[104] and many faradaic reactions of interest involve consumption of the electrolyte.[109] To increase meniscus stability, Stoerzinger et al. suggested addition of non-interacting salts to the electrolyte, with appreciable success.[28] Third, the procedure relies on wettable surfaces. That means that for a water-based electrolytes only hydrophilic samples can be investigated, posing restrictions on the sample and electrolyte choices.[26] Lastly, X-ray damage of the solid or radicals created during water radiolysis are of concern, as in all X-ray based techniques.[27,110] The severity of X-ray damage strongly depends on the cell design and beam energy, intensity and size,[111] complicating analysis of the solid-liquid interface at modern high-flux beamlines.

### B. Membrane XPS

As described in section 2.2, membrane-based operando cells are now available with exceedingly thin solid membranes, which thus become photoelectron transparent, for example in the holey $SiN_x$ cell and the ion exchange membrane cell geometry (Figure 3b-c). They are thus also suitable for operando XPS measurements. Compared to meniscus XPS, these cells offer the advantage of decreasing the effect of beam damage because they allow replenishing of the electrolyte. Yet they also suffer from more convoluted spectroscopic data because the interface is now buried by a solid membrane. Depending on the scientific question and material characteristics, one or the other geometry will thus be more promising.

## 3.3 Interface sensitivity examples
### A. Surface composition and active sites

Generally, operando XPS can reveal chemical composition, the oxidation state and built-in electrical potentials of both the liquid and the solid. Both membrane and meniscus XPS have been used to reveal the active (surface) phases of various electrochemical materials under operating conditions. This information is needed to understand the reaction mechanisms and in turn identify predictive design rules for materials optimization based on true active surface structure – activity relationships. For example, operando XPS confirmed the active phase of Co-metal based water splitting electrocatalysts in alkaline media. At oxidative potentials (i.e., under reaction conditions for the OER), complex $Co(OH)_2$ and CoOOH phases evolve at the solid-liquid interface, indicating that cobalt oxyhydroxide is the active phase (Figure 6a and b).[112] Experimentally, the phases were assessed through the peak shape, binding energy position of the Co 2$p$ peak and relative peak intensities. All of these change reversibly with applied potential, with a reduction of the surface hydroxide layer back to Co metal at negative potentials. The shift of the liquid and gas phase $H_2O$ O 1$s$ contribution with applied potential shown in Figure 6b further confirms that the desired potential was applied at the measurement spot. The use of XPS was essential for the material-specific discovery, because the active phase only exists in the top few nanometers,[113] or ultimately in only atomic-layer thicknesses.[14] Similar observation of transition metal (oxy)hydroxides were found for various transition metal based electrocatalysts[113–115] under reaction conditions. For example, even thin-film perovskite oxides like $La_{0.2}Sr_{0.8}CoO_{3-\delta}$ form a surface layer of Co oxyhydroxide.[116] Similarly, Pt electrodes for water electrolysis were investigated both as cathode[117] and anode,[108] revealing reversible surface oxidation, and formation and dissolution of CuO was found to dictate the activity in oxide-derived Cu electrodes for $CO_2$ reduction reaction.[118] In other words, XPS reveals insights about the surface chemistry far beyond traditional (surface) Pourbaix diagrams. In combination with operando XAS, it was even possible to identify the chemical nature of the active oxygen sites during deprotonation of hydroxyl groups, a key step in oxygen evolution reaction in state-of-the art OER catalyst $IrO_2$.[119]

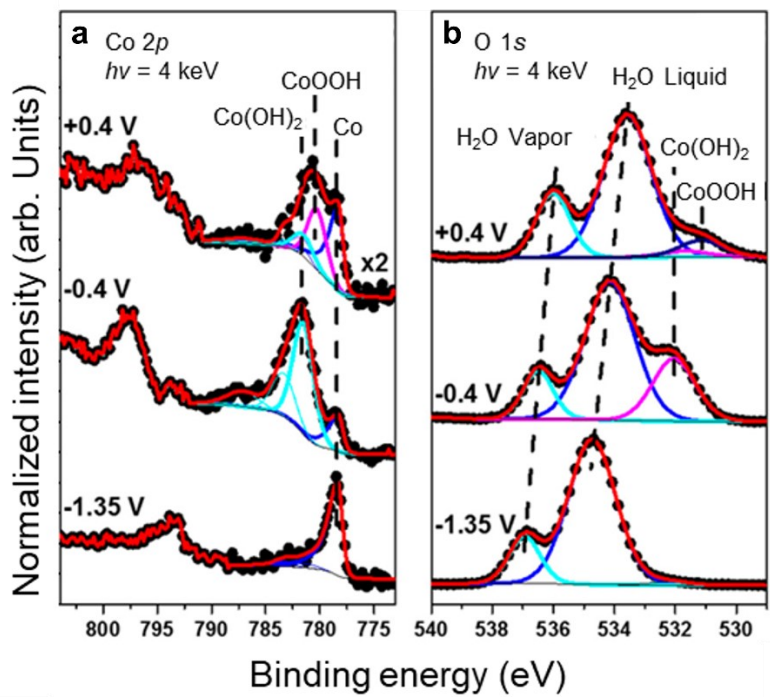

**Figure 6:** (a) Co 2p, and (b) O 1s, meniscus photoelectron spectra of a Co metal foil in 0.1 M KOH, collected at −1.35, −0.4, and +0.4 V versus Ag/AgCl at hυ = 4 keV. Changes in Co 2p peak shape indicate the formation of Co hydroxides and Co oxyhydroxides. The O 1s spectrum contains components from liquid and gas-phase water alongside species from the electrocatalyst. The peak shift of the $H_2O$ peaks confirms the potential drop due to the applied bias. Reprinted (adapted) with permission from Han, Y. et al., *The Journal of Physical Chemistry B* 2018, *122* (2), 666–671. Copyright 2018 American Chemical Society.

B. **Potential profile**

The binding energy position can also vary due to electrostatic potentials at interfaces or surfaces. In this case, a "rigid" shift of all core levels is expected, i.e. all core levels of the same compound shift by the same amount, in contrast to the "chemical shift" where cations and anions display a shift in opposite directions upon bond formation. Generally, charge accumulation and band bending is observed at interfaces and surfaces (both in the solid and the liquid). When potential gradients exist, photoelectrons emitted in each atomic layer are subject to slightly different potential and thus appear at shifted binding energies.[47] Combined with the depth-dependent intensity attenuation (equation 3), this situation leads to asymmetric broadening of the XPS peaks (Figure 7a). Assessing this broadening through a layer-resolved deconvolution taking into account the attenuation, or a simple estimate of the peak broadness through the full width at half maximum can therefore be used to extract the potential profile across the interface. This results in an experimental observation of the electrochemical double layer forming at electrode-electrolyte interfaces (Figure 7a),[5] and the band alignment of the photocatalyst in photoelectrochemical cells.[120,121] The FWHM trend of the core-level peaks of electrolyte and electrode can be investigated simultaneously, resulting in a direct probe of the potential distribution in the liquid and the solid at applied potential, allowing experimental verification of classic and state-of-the-art models of the electrochemical double layer, which is essential for the understanding of virtually all electrochemical electrode processes. We note that this is only experimentally accessible for low electrolyte concentrations, where the electrochemical double layer thickness extends far enough into the electrolyte to contribute substantially to the total photoelectron intensity, which decays exponentially from the solid-gas to the solid-liquid interface (equation 3). For higher electrolyte concentrations or smaller potential steps at the solid-liquid interface (Figure 7b), the components originating from the "bulk" part of the liquid layer have essentially zero shift in binding energy due to a flat potential profile. In this case, a strong dependence of the potential on the peak position is only observed at the solid-liquid interface, where the relative photoelectron intensities are negligible compared to the region near the solid-gas interface.

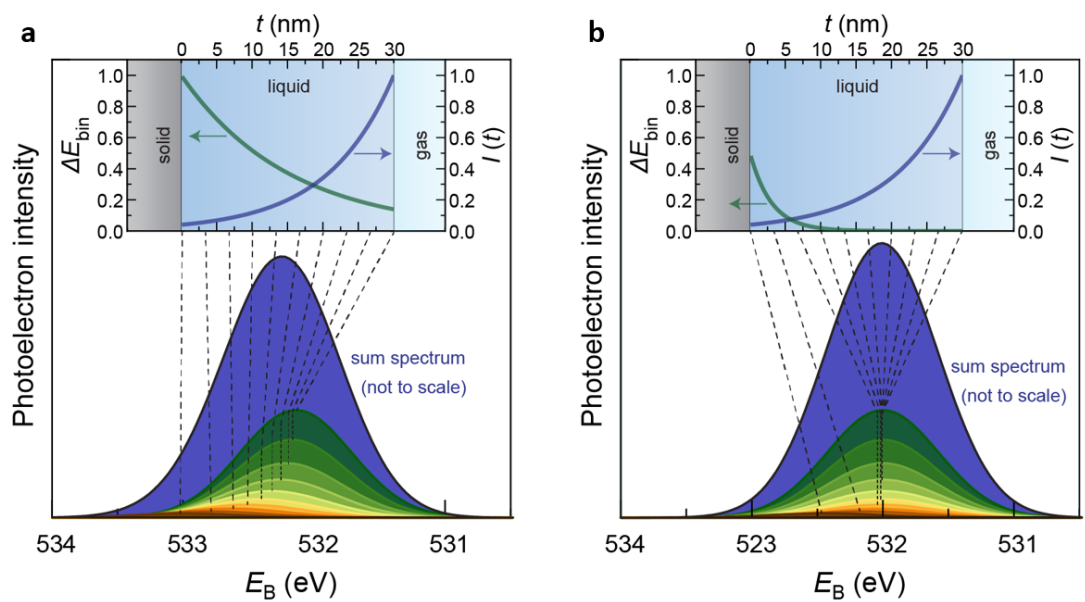

**Figure 7:** (a) Spectral broadening of the O1s core level of the elements belonging to the liquid phase, for a positive potential (1V) applied to a metal electrode with a dilute 30 nm aqueous electrolyte (0.4 mM KOH). The colored spectra indicate representative spectra from a 0.3 nm slice of the liquid phase, with position and spectral contribution simulated according to the potential profile and intensity profile from the inset (according to equation 3). The binding energy shift is the inverse of the potential profile, which we assumed based on the Gouy-Chapman model for the double-layer potential in comparably dilute electrolytes using the values from reference [5]. The blue spectrum is the sum over all individual spectra, not drawn to scale. (b) Spectral broadening for a smaller surface potential (0.5 V) and more concentrated electrolyte (10 mM KOH). The broadening is hardly visible because only the slice-spectra near the solid/liquid interface (i.e. the spectra with the smallest relative intensity) are shifted appreciably. This is a result of the small double layer thickness for higher electrolyte concentrations. Insets show the binding energy shift resulting from the potential profile and the photoelectron intensity decay in the electrolyte layer (equation 3). Dashed lines indicate the peak corresponding to the relative position in the electrolyte layer. The difference in peak shape resulting from the differences in double layer can be qualitatively seen by the different maximum intensity of the sum spectra in a and b, while they have identical integrated intensity.

## 3.4 Need for simulations

The understanding of XPS data is most robustly done through a comparison to experimental reference spectra from seminal works as presented in reference.[122] The reason lies in the complex peak shapes for many elements of interest. For example, multiplet splitting occurs when the binding energy depends on exchange interaction between the remaining core level electrons (after photoionization) and unpaired electrons in the valence band, and is relevant in many transition metal based compounds. The ground state can further consist of multiple states including core holes. Taking CoO as an example, the ground state wave function is a superposition of multiple states of $d$-orbital occupation, which includes holes $\underline{L}$ in the O $2p$ orbital, mathematically expressed as $\Psi = a|d^7\rangle + b|d^8\underline{L}\rangle + c|d^9\underline{L}^2\rangle$ with coefficients $a + b + c = 1$ and the number $i$ of electrons in the $d$-orbital $d^i$.[123] During the photoemission process, this leads to differences in the final state energies, too, implying that the spectral shape cannot be well represented with simple and empirical peak shapes available in most XPS analysis software tools. Instead, recent efforts going beyond comparison to reference spectra rely on DFT calculations[116,124,125] or so-called cluster calculations for binding energy comparison to experimental spectra. Cluster calculations are ab initio calculations of the electronic structure capable of describing the multiplet ligand-field theory, while reducing computational time through restriction of the many-electron effects to a single transition metal site plus its ligand neighbors.[126] In addition to such simulations of the chemical properties, spectra can also be simulated to include the potential profile observed at the solid-liquid interface.[5] To assess the thickness of a thin electrolyte layer and to predict expected intensities for a given solid-liquid interface, simulations can be performed using the SESSA software package and database (Simulation of Electron Spectra for Surface Analysis) developed by NIST.[127] We recommend to perform SESSA simulations of a planned sample and cell geometry before a new operando XPS experiment to judge if the relevant core levels can exhibit suitable intensity or if geometry optimization is necessary.

## 3.5 Future developments and interface-sensitivity of meniscus XPS

For operando XPS and XAS cells, the technological advancements around the world are currently accelerating, because of both the promises and challenges of the technique. Investigation of the solid-liquid interface is now becoming possible in laboratory-based APXPS systems,[64,107,128] where the experimental turnaround for a specific experiment can be faster, and the issues of beam damage might be less severe (at the high cost of severely prolonged integration times). A key point for improvement is the stability of the meniscus, for example using a fine capillary close to the region of interest to balance the evaporation rate or sample cooling,[106] perhaps in parallel with the chemical strategies to stabilize the liquid layer.[28]

An essential point for the interface-sensitivity of operando XPS is the suitable selection of the X-ray energies. The very first report on meniscus XPS already included SESSA simulations.[25] It was found that "tender" X-rays with energies of around 4 keV optimize the signal intensity of a thin overlayer on a chemically different substrate (in this case 1 nm Fe on a Si substrate) through a water meniscus.[25] Later analysis showed that the ideal energy for the detection of species of the liquid side of the solid-liquid interface is also in the tender X-ray regime[25,37] and that the ideal energy also depends on the selected core level binding energy.[26] A similar analysis was also performed for illumination and detection through a thin membrane.[129]

For our electrocatalyst example from section 2.2, the situation is very different compared to the previous discussions in refs. [25,26]: not only the total signal of the interfacial layer but also the interface-sensitivity need to be considered. This is especially important for a thin surface layer containing the same elements as the underlying bulk of the solid. The analysis below will show that in this case hard X-rays (which are of course favorable for photoelectron penetration of the meniscus) lead to an overshadowing of the interface information by the electrons emitted from the subsurface of the solid. As electrochemical and other reactions occur at this interface and depend critically on the surface structure and chemistry, the use of such X-rays therefore impedes obtaining the relevant information, making the correct choice of the X-ray energies even more important for such systems.

To assess this situation quantitatively, SESSA simulations were performed on a system resembling the experimental setup in meniscus XPS with a $LaCoO_3$ electrode, chosen as a typical representative of a perovskite electrocatalyst without easily dissolvable species.[130–132] Inspired by our findings for $LaNiO_3$,[14] it is assumed that the Co chemistry of the top one to two unit cells changes as a function of applied potential. We thus divide the 20 nm electrode into a 0.8 nm $LaCoO_3$* surface layer and a $LaCoO_3$ bulk layer (Figure 8a). Four representative X-ray energies were chosen, which can be obtained with X-ray sources available for laboratory-based experiments.[133] The simulated spectra are shown in Figure 8b with insets for the Co 2$p$ and Co 3$p$ peaks. The simulation reveals that the total Co peak intensities increase with increasing photon energies, while the O 1$s$ peak intensity decreases with increasing photon energies. To consider the interface-sensitivity, the Co peak intensities are divided into a surface and a bulk species. Figure 8d shows that the signal from the surface layer is maximized for the "tender" X-rays (2697 eV). However, the relative contribution of the surface layer shows a fundamentally different behavior: the relative surface intensity decreases monotonically with increasing photon energy (Figure 8d), because of the inelastic mean free path increase. The key outcome of this analysis therefore is that the *relative intensity* of the surface contribution scales differently with photon energy than the *total intensity*. These simulations match experimental photon-energy-dependent measurements qualitatively (Figure 8c).[107] To track small spectral changes originating from a thin surface layer, one should therefore choose the minimum photon energy that still yields sufficient total count rate.[133] Ideally, multiple X-ray sources (or various X-ray energies in a synchrotron experiment) should be used to carefully extract the surface information through comparison of the respective contribution to the total signal.

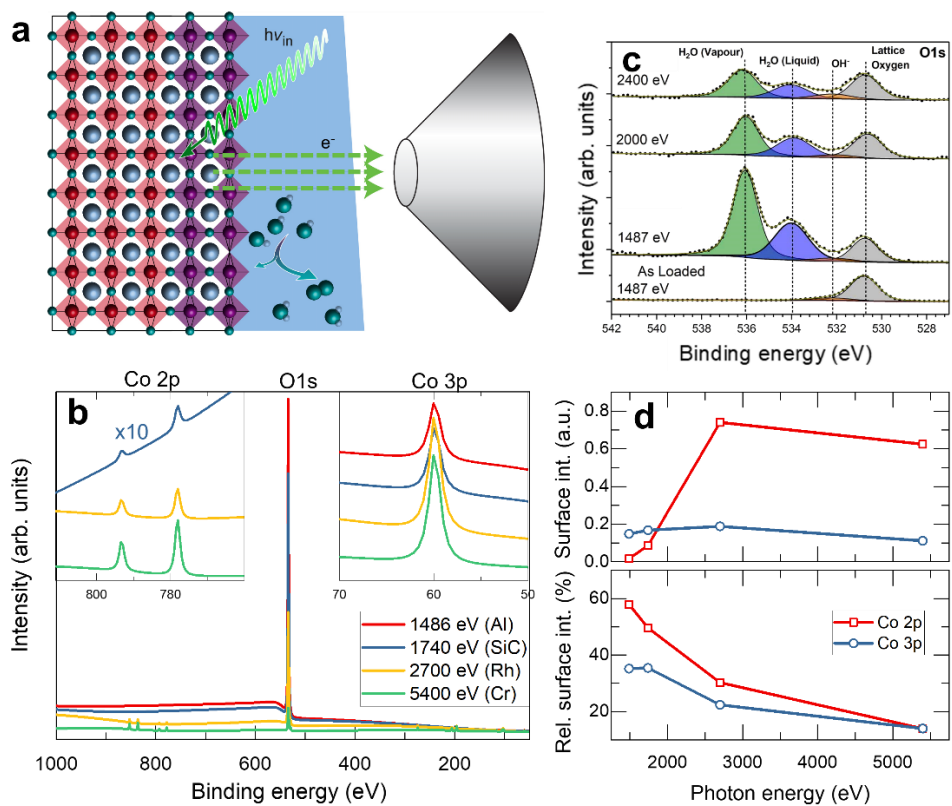

**Figure 8:** (a) Schematic for SESSA simulations (not drawn to scale). A 0.8 nm LaCoO$_3$* surface layer (violet) on a 19.2 nm LaCoO$_3$ OER electrocatalyst (red) is interfaced with a 20 nm meniscus of liquid water. (b) Simulated spectra. The spectra are dominated by the O 1$s$ peak, which stems mostly from the liquid layer. The insets show a zoom-in to the Co 2$p$ and Co 3$p$ core levels (left and right inset, respectively). Spectra in the left inset are offset vertically for easier comparison (1486 eV is off the scale), and spectra in the right inset are displayed to-scale. (c) O 1$s$ spectra of a TiO$_2$/H$_2$O interface at different photon energies compared to a pristine TiO$_2$ surface. All spectra are normalized to the lattice oxygen peak. Reprinted (adapted) with permission from Byrne, C. *et al., J Phys D Appl Phys* 2021, *54* (19), 194001. (d) Top: Simulated intensity of the Co2$p$ and Co 3$p$ core levels, considering only photoelectrons from the two-unit-cell surface layer. Bottom: Relative contribution of the surface signal to the total intensity. (a), (b), (d) reprinted from Baeumer, C. Operando Characterization of Interfacial Charge Transfer Processes. *J. Appl. Phys.* 2021, *129* (17), 170901. https://doi.org/10.1063/5.0046142, with the permission of AIP Publishing.

Another possibility to achieve the desired interface sensitivity is turning to angle-dependent measurements.[134–137] For the solid-liquid interface, however, it remains a formidable challenge to vary the photoemission angle in the required range of ~40° necessary to extract a depth profile while maintaining a stable meniscus. A more promising route for ultimate interface-sensitivity in XAS and XPS might be the small angle variation (a fraction of a degree up to ~3°) necessary to achieve depth profiling in the so-called standing wave or near-total-reflection approaches. X-ray standing waves, i.e. periodic fields of antinodes and nodes can be generated at a superlattice of optically dissimilar materials by interference of incoming and reflected photons. The vertical position of antinodes of the standing waves shifts with varying the incident angle, resulting in precise depth-selective photoemission intensity modulations. As a result, sub-unit cell depth resolution can be achieved.[33,138] For our example of a surface transformation in perovskite oxide electrocatalysts, ex situ standing wave XPS already demonstrated that the chemical changes were confined to a 4 Å surface layer.[14] Few pioneering studies by the groups of Fadley, Bluhm and Nemšák applied standing wave XPS to solid-liquid interfaces and revealed great potential.[30,31] However, long integration times are needed because all relevant core levels must be measured at a multitude of angles to extract the depth-profile of chemical states and composition from the angle-dependent intensity modulation. Recently, we introduced near-total-reflection XPS as an alternative approach without the need for the accurate preparation of multilayer mirrors and detailed knowledge about layer thicknesses and their optical properties. This methodology makes use of X-ray optical effects in the near total reflection (NTR), where the penetration depth of X-rays, and therefore the locus of photoelectron generation, can be tuned via X-ray energy and incidence angle, resulting in a depth-resolution of ~1 nm.[139]

# 4 Surface X-ray diffraction – Crystal Truncation Rod

## 4.1 Theory of the technique

A crystalline material's atomic structure can be determined by measuring the diffraction of X-rays by the crystals (Figure 9a), as shown by von Laue in 1912.[55,140] Let us consider a crystal whose lattice can be indexed using the orthogonal lattice vectors: $\{\vec{a}, \vec{b}, \vec{c}\}$. The scattering intensity from each atom within the unit cell is given by the structure factor, $F(\vec{q})$

$$F(\vec{q}) = \sum_j f_j e^{i\vec{q}\cdot\vec{r_j}} \qquad 5$$

where $f_j$ is the form factor for an atom located at position $r_j$ in the unit cell and $q$ is the scattering vector (the difference between the diffracted and incident wave vectors). The form factor $f_i$ depends on the atomic number of the scattered atom, and therefore, this technique cannot be used to either structurally refine the position of light elements, or to distinguish between elements that are close together in the periodic table.

In order to obtain the scattering intensity from a 3D crystal, the contribution of all unit cells in the crystal needs to be considered:

$$I = \left| F(\vec{q}) \sum_n e^{i\vec{q}\cdot\vec{R_n}} \right|^2 \qquad 6$$

where $\vec{R} = n_1\vec{a} + n_2\vec{b} + n_3\vec{c}$ describes the origin of each unit cell in the lattice. Therefore, any change in atomic positions will be reflected as a change in the scattered intensity. In the simplest case, for an ideal bulk material that extends infinitely in all three directions, the values of $n_1$, $n_2$ and $n_3$ range from -∞ to +∞. This results in a diffraction pattern with a set of distinct delta functions at defined values of $q$:

$$I \propto \left| F(\vec{q}) \sum_{h,k,l} \delta(q_a a - 2\pi h)\delta(q_b b - 2\pi k)\delta(q_c c - 2\pi l) \right|^2 \qquad 7$$

When the following conditions are met: $\vec{q}_a\vec{a} = 2\pi h, \vec{q}_b\vec{b} = 2\pi k, \vec{q}_c\vec{c} = 2\pi l$, large scattering intensity is observed, known as Bragg peaks (Figure 9b); $h$, $k$ and $l$ are the Miller indices.

For the more relevant case of semi-infinite crystals, i.e. 3D crystals that have been terminated to form a surface in the $c$ direction, the value of $n_3$ ranges from 0 to ∞. Mathematically, by solving equation 6 for this specific boundary condition, the resultant intensity of the diffraction pattern is given by:

$$I \propto \left| F(\vec{q}) \sum_{h,k} \delta(q_a a - 2\pi h)\delta(q_b b - 2\pi k) \frac{1}{\sin\left(\frac{q_c c}{2}\right)} \right|^2 \qquad 8$$

Therefore, the diffraction pattern for semi-infinite crystals also includes streaks of intensity between the Bragg peaks called crystal truncation rods (CTR). Figure 9b shows the ideal diffraction pattern for a 3D infinite crystal and a semi-infinite crystal. The non-zero scattering intensity between the Bragg peaks over a range of q is one of the most important characteristics of CTRs. From this extended scattering, one can determine the structure factor and hence the atomic positions of the surface or interface layer. CTRs – also referred to as surface X-ray diffraction – have been applied to study roughening,[141] surface reconstructions,[142,143] local adatoms,[144] buried interfaces between two well-defined crystals,[145,146] solid-liquid[8,147] and solid-gas[148] interfaces. One can show that these changes in scattered intensity are most prominent at the anti-Bragg (the region between two Bragg peaks) of the CTRs. Consequently, the relevant structural parameters can be determined by measuring a range of CTRs.

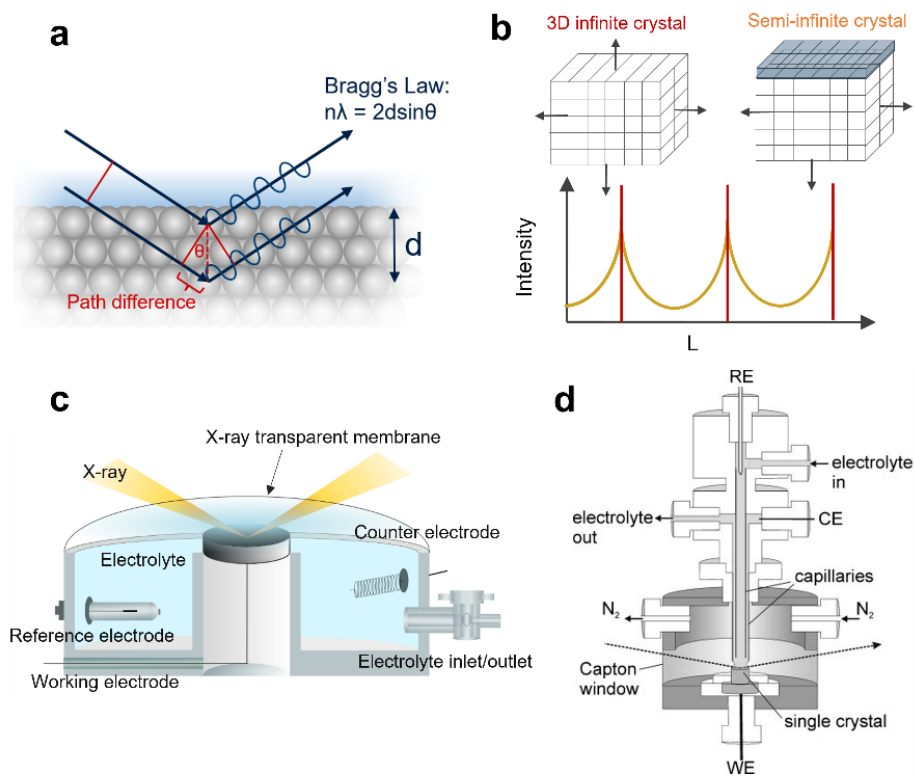

**Figure 9:** (a) Scattering of X-rays from a periodic crystal lattice. (b) Ideal diffraction intensities as a function of the reciprocal lattice unit for a 3D infinite crystal (red) and a terminated surface (orange). (c) Schematic of an X-ray thin layer cell[54] and (d) Droplet cell. Reprinted from Electrochim Acta 53, Magnussen, O. M., Krug, K., Ayyad, A. H. & Stettner, J., In situ diffraction studies of electrode surface structure during gold electrodeposition, 3449–3458. Copyright (2008) with permission from Elsevier.[149]

## 4.2 Experimental measurement

While CTRs are extremely sensitive to changes in the surface atomic structure, the intensities measured in the anti Bragg region are several orders of magnitude lower than the bulk Bragg reflections. Therefore, synchrotron light sources, and particularly beamlines with high photon flux are required to measure these low intensities with good signal to noise ration. CTR measurements can be performed at ambient conditions, at a range of temperatures, for samples placed in controlled gas or liquid atmospheres, and are thus ideal to probe the electrochemical interface. Layer-by-layer atomic information can be obtained, down to a resolution of 1-10 pm in the surface normal direction.[150–153] However, this resolution is not always realized in practice owing to limitations of instrumental resolution, quality of crystalline surface and structural refinement data analysis. In fact, CTR measurements rely heavily on atomically-smooth single crystal or epitaxial thin film surfaces. Surface roughness of even a few nm can result in a sharp decrease in the intensity at the anti-Bragg positions, resulting in poor structural refinement.[141]

Typical cell designs include an X-ray thin layer cell[53,54,153] or a droplet cell.[52,154] In an X-ray thin layer cell (Figure 9c), a very thin layer of electrolyte is trapped between the sample surface and a Kapton film. The film is attached to the cell with an O-ring. On the one hand, a continuous layer of electrolyte should be present on the sample to enable mass transport, but on the other hand, the thickness of the electrolyte layer should not significantly attenuate the measured intensity. The cell body is made of a chemically inert material such as Kel-F and the counter and reference electrodes are held in place by Kel-F clamps. The catalyst surface protrudes slightly higher than the body of the cell, allowing for grazing incidence measurements. Another widely used cell set up is the droplet cell configuration, Figure 9d. Here, a small drop of electrolyte is ejected from a capillary and is in contact with the sample surface. The volume of the electrolyte can be controlled by a syringe. This configuration ensures that the electrolyte is only in contact with the sample surface, however, droplet stability, specifically under gas evolution conditions, can result in significant issues.

Once assembled, the cell is mounted on a four or six circle goniometer, which allows for multiple degrees of freedom of motion of the sample with respect to the incoming beam. An orientation matrix, defining the relationship between the goniometer angles and the scattering vector, is constructed to measure the diffracted intensity at different points in the reciprocal space. For non-specular CTRs, i.e. those with an in-plane component of the momentum transfer, the incident angle is fixed, so that the penetration depth of the X-rays and the size and shape of the beam footprint remain constant. The size of a focused X-ray beam is usually in the range of a few hundred micrometers. Therefore, the information provided by this technique is spatially resolved over this length scale. The number of inequivalent CTRs that need to be measured is determined by the symmetry of the bulk crystal structure. The CTRs are measured using pixel-resolved area detectors, which can simultaneously measure the signal and background. Depending on the signal to noise ratio, it may take several hours to record one CTR, and therefore the stability of the sample under potential control is also critical.

## 4.3 Interface sensitivity examples

**A. Surface structure**

Surface diffraction has been widely employed to detect changes in surface structure as a function of applied potential. For example, it has provided unique insights into the degradation of Pt-based catalysts for application in fuel cells. During electrochemical oxidation of Pt(111) single crystal surfaces, fitted CTR surface structures have shown that Pt surface atoms leave their lattice sites, thus inducing surface restructuring via a place exchange mechanism.[51,147,155–158] Recent studies by Magnussen *et al.*[159] also used CTR to directly compare the atomic-scale surface oxidation and degradation mechanisms on Pt(111) to Pt(100) surfaces. Interestingly, based on fitting changes in their CTR data, they found that on applying oxidizing potentials to Pt(111), the Pt place exchange is triggered. The Pt remains above its original site when the upper potential values are lower than 1.15 V, and this process is reversible since this Pt can return back to its original site. However, in the case of the Pt(100) surface, the CTR data (Figure **10**a) can only be modelled using a surface structure that suggests the Pt are vertical displaced by $d_{ex}$<1.42 A and are laterally displaced to a bridge site (Figure **10**b, c). Such a lateral movement of Pt atoms during the place exchange mechanisms results in the immediate removal of a second atom and instability of the surface. Through this detailed structural analysis of the surface, the authors were able to unravel the physical origin of the higher corrosion of Pt(100) compared to Pt(111). Similarly, Over et al.[160] have used surface scattering to determine the degree of iridium corrosion in model electrodes of 50 Å single crystalline $IrO_2$(110) in acidic electrolyte supported on slightly bulk-reduced $TiO_2$(110). Notably, in this case, the absence of any significant changes in the CTRs over ~26 hours at 50 mA/cm$^2$ suggests that the $IrO_2$(110) is stable under these conditions, corresponding to a maximum of 0.10 monolayer corrosion. Therefore, in this case, CTR measurements reveal the remarkable stability of $IrO_2$(110) under anodic oxygen evolution reaction conditions.

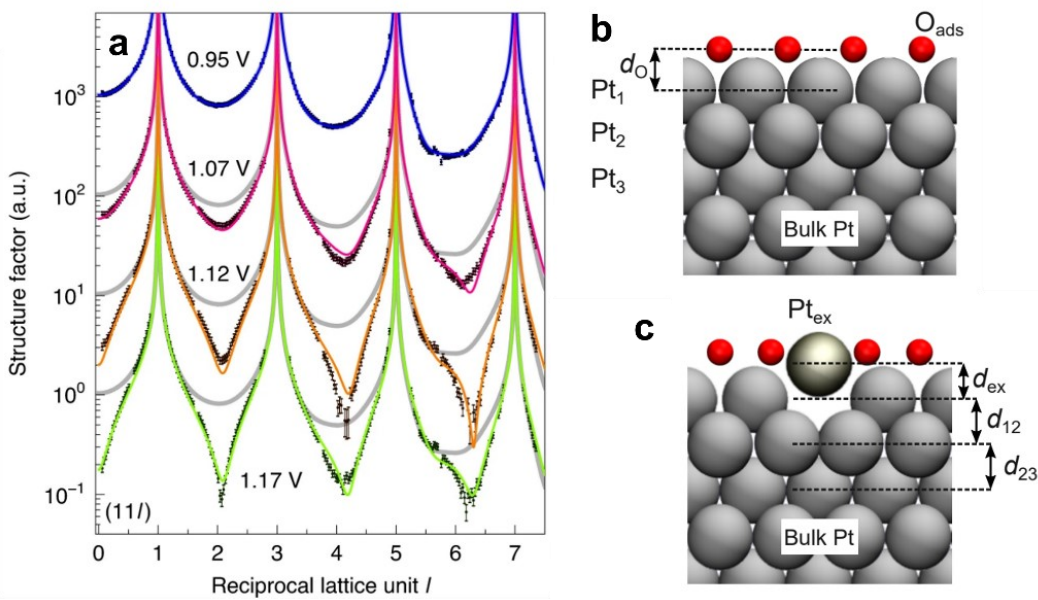

Figure 10: (a) (11L) rod of Pt(100) measured at 68 keV at potentials 0.95 $V_{RHE}$, 1.07 $V_{RHE}$, 1.12 $V_{RHE}$ and 1.17 $V_{RHE}$ in 0.1 M $HClO_4$. In addition to the experimental data (black dots), which are offset by a factor of ten with respect to each other, the best fits (coloured lines) and the CTR fits (grey line) for the smooth surface at 0.95 V are shown. The decrease in intensity at the anti-Bragg positions at potentials higher than 1 $V_{RHE}$ is indicative of the place exchange of surface platinum atoms. Schematic of the CTR fitting model for the Pt(100) surface at (b) 0.95 $V_{RHE}$ and (C) 1.07 $V_{RHE}$, 1.12 $V_{RHE}$, 1.17 $V_{RHE}$. The fit parameters for this analysis include the interplanar distances, $d_{ex}$, $d_{12}$, $d_{23}$, coverage of exchanged platinum $Pt_{ex}$ and Debye Waller factors for atoms $Pt_i$. Reprinted by permission from Springer Nature: Nature Catalysis, Structure dependency of the atomic-scale mechanisms of platinum electro-oxidation and dissolution, Fuchs et al. Copyright 2020.

**B. Surface adsorbates**

In principle, CTR can also be used to determine the position and coverage of surface adsorbates as a function of potential. However, this is particularly challenging for small molecule catalysis, since the adsorbed atoms tend to have a significantly smaller scattering factor compared to the bulk metal atoms, owing to their smaller atomic number. However, this issue can be circumvented in certain cases. For example, in a $RuO_2$(110) unit cell, there are two Ru atoms at coordinates of (0,0,0) and (0.5, 0.5, 0). On solving Equation 5 for this specific unit cell, it becomes apparent that for h+k values that are odd, the scattering contribution of the two independent Ru atoms vanishes. Since h+k=odd rods do not, in principle, have any Ru contribution, they are known as the 'oxygen rods'.[53,54,161] Shao-Horn et al. have use this to study the surface redox reactions and OER mechanism on $RuO_2$ single crystal surfaces.[53,54] In this study,[54] they find that the oxygen rods, (10L) and (01L), on $RuO_2$ (110) single crystal surfaces show significant changes in the anti-Bragg region with increasing potential from 0.5 $V_{RHE}$ to 1.5 $V_{RHE}$ (Figure 11a, b). These changes can be fit to surface structures where the bond distance between the adsorbed O atom on surface Ru coordinatively unsaturated (CUS) and Ru bridge sites decreases with increasing potential from 0.5 $V_{RHE}$ to 1.5 $V_{RHE}$. Most significantly, at OER relevant potentials, the authors found the presence of an -OO intermediate on the Ru CUS site. By combining the experimental observations with DFT analysis, they concluded that this -OO species was stabilized by neighboring -OH species on a Ru bridge or Ru CUS sites. This work thus revealed the presence of a new OER pathway involving the final deprotonation as the rate-determining step for the reaction. In a follow-up study,[53] they extended this analysis to the (100) and (101) single crystal facets of $RuO_2$, where the binding energy of oxygen on the Ru CUS site decreases from the (110) to the (100) and (101) surfaces. While they found a similar OER mechanism on the (100) surface, the higher OER activity observed was explained by decrease in the binding energy of the -OO species on the surface. For the (101) surface, a completely oxidized surface was found at high potentials, and the activity was found to decrease relative to the (100) surface. This was attributed to a change in the rate-determining step to O-O bond formation. By progressively weakening the binding energetics, the authors were not only able to measure the theoretically predicted volcano relationship between activity and binding energetics but were also able to detect a change in the rate determining step for OER. Therefore, although the direct determination of adsorbates is challenging for CTR measurements, they provide unprecedented insights into the active sites and reaction mechanisms for electrochemical processes.

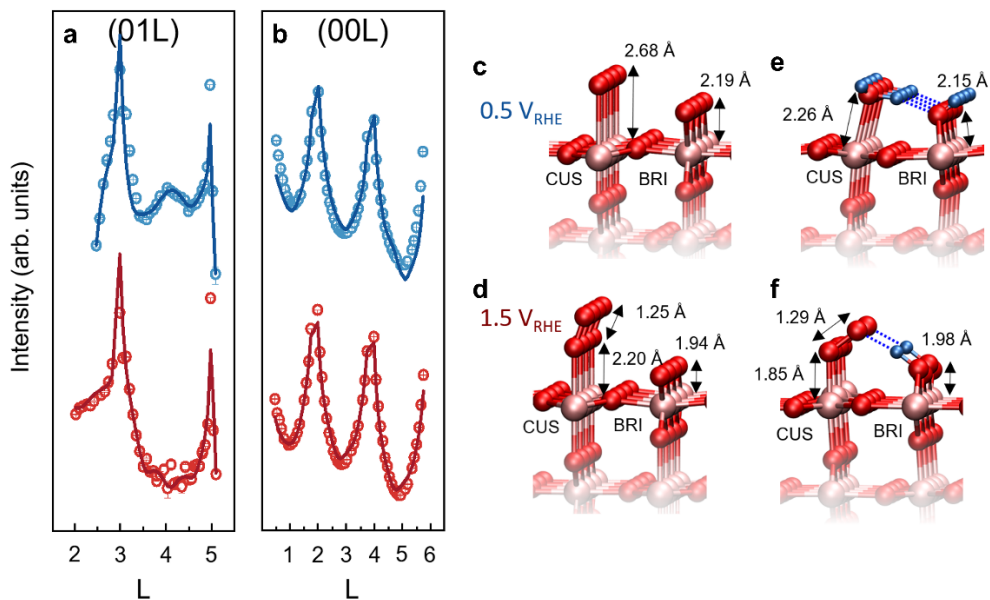

**Figure 11:** (a) (01L) and (b) (00L) rods measured at two potentials, 0.5 $V_{RHE}$ (blue) and 1.5 $V_{RHE}$ (red). Open points denote the experimentally measured intensities and the best-fit results from the fitting process are shown as solid lines of the corresponding color. Ball and stick models for the best-fit structures obtained for the (c) 0.5 $V_{RHE}$ and (d) 1.5 $V_{RHE}$ crystal truncation rod data and the corresponding most stable adsorbate configuration at (e) 0.5 $V_{RHE}$ and (f) 1.5 $V_{RHE}$ obtained using DFT. All measurements were made in 0.1 M $HClO_4$. Pink, red and blue spheres represent Ru, O and H atoms respectively. Reproduced from Ref. [54] with permission from the Royal Society of Chemistry.

## C. Interfacial Electrolyte Structure

Ion adsorption at solid-liquid interfaces is critical to a number of environmental and chemical processes such as growth and dissolution of minerals in aqueous environments,[162,163] electrokinetic phenomena in colloidal science[164] and geochemical applications.[165] From an electrochemical perspective, hydrated cations can also interact with adsorbed intermediates and change the kinetics of the reaction, either by (de)stabilizing the intermediates[166] or by blocking the actives sites on the surface.[167,168] Furthermore, ordering of water molecules at the interface can also change the reaction kinetics, as has been recently demonstrated on Pt surfaces[169,170]. Therefore, probing electrolyte structure is becoming increasingly important to gain a holistic understand of solid-liquid interfaces. Crystal truncation rod analysis has been successfully applied to study adsorption of heavy ions such as $Rb^+$ and $Sr^{2+}$ at $TiO_2(110)$ surfaces.[171,172] The CTR data obtained was fit to a structure where the ions are present at tetradentate sites, located between the Ti bridge and Ti CUS sites, at a distance of ~3.5 Å above the surface Ti-O plane. The effects of electrolyte structure on the CTR intensity can be subtle. Recent work by Kawaguchi et al. has proposed a method to normalize the out-of-plane (00L) rods to visualize changes more easily, and use a direct inversion method to determine the electron density as a function of distance into the liquid layer.[173] This method has thus enabled the elucidation of $Cs^+$ ordering at the Pt(111)[174], $RuO_2$(110) and $RuO_2$(100) surfaces[175] at a vertical distance of ~3.5 - 4 Å from the metal centers in 0.1 M CsF, with the degree of ordering increasing with lowering the potential.

In addition to cation adsorption, ordering of water at metal-electrolyte[8] and oxide-electrolyte[176,177] interfaces can be different from bulk water. As discussed above, XAS can be used to determine differences in the hydrogen bond environment of water at interfaces. Complementary structural information about the position of these water molecules can be obtained using CTR. For example, Hussain et al. [177] have demonstrated that the ordering and position of water molecules near the surface of $TiO_2$(110) can be determined using CTR. They found that their CTR data could be best-fit to a model showing the presence of an ordered water layer at tetradentate sites and a distance of ~3.8 A from the surface Ti-O plane; these results were corroborated with molecular dynamics simulations.[177] Similarly, an ordered layer of water molecules was detected at ~3.75 Å from the surface Ru-O planes by fitting normalized (00L) CTR data, on $RuO_2$(110) surfaces. While the degree of ordering decreased from 0.75 $V_{RHE}$ to 1.50 $V_{RHE}$ in 0.1 M LiOH, 0.1 M NaOH and 0.1 M KOH, it was found to be invariant on the nature of the cation.[176] These studies mark an exciting area for further research, considering the uniqueness of this technique in mapping the structure and ordering of the electrolyte at the interface.

## 4.4  Need for simulations

Existing tools such as ROD[178] and GenX[179] can be used to refine surface X-ray diffraction data. However, physical insight into the surface structure provides a useful starting point for further optimization and ensures that fitting algorithms converge to the best solution. Therefore, combining surface diffraction measurements with theoretical calculations is very useful. Specifically, for electrocatalysis, the position of several small atoms such as H, C, N cannot be resolved with surface diffraction data alone.[8,55] For example, although recent studies have demonstrated a shortening of the Ru-O bond length with increasing potential from 0.5 $V_{RHE}$ to 1.5 $V_{RHE}$, the exact nature of the adsorbed intermediates is unknown.[53,54] One approach would be to compare these bond lengths to reference literature of Ru-O bond lengths for adsorbed *$H_2O$, *OH, *O and *OOH species. However, such reference data can be difficult to obtain. In this example, comparison of the experimentally observed bond lengths with DFT predictions for the most stable surface structure as a function of potential was shown to be extremely useful for assigning the nature of surface adsorbates as a function of potential.

With advances in instrumentation and measurement techniques leading to faster data acquisition[152], data fitting is expected to become the main bottleneck for this technique. Therefore, increasing efforts are also being parallelly invested in tools that can simulated surface diffraction spectra[180] as well as machine learning approaches that can facilitate model development and data fitting[181]. For example, an ab initio tool that can provide simulations of surface resonant X-ray diffraction (i.e. surface X-ray diffraction measurements performed in an energy range that corresponds to the absorption edge) has been recently developed.[180] Here, DFT is used to determine the resonant scattering factor of an atom, which serves as an input to calculate the diffraction intensities. Interestingly, a numerical diffractometer that accounts for factors such as beam polarization, incident and exit angles has also been implemented in this simulation. This tool has been already applied to investigate two very different systems - magnetite thin films deposited on silver as well as bromine atoms on copper. Such theoretical tools are expected to significantly increase the speed and accuracy of data fitting and interpretation.

# 5   Conclusion and outlook:

In this chapter, we summarized the state of the art in operando XAS, XPS and CTR characterization of the solid-liquid interface with examples from electrocatalyst research. Undoubtedly, the increasing commercial availability, accessibility and continuous efforts to develop beyond-state-of-the art instrumentation at synchrotron facilities and in laboratories around the world will make the operando characterization of solid-liquid interfaces invaluable for research activities in a variety of fields like electrochemical solid-liquid interfaces for energy technologies, including batteries.

Remaining challenges are related to combining interface-sensitivity with resolution in additional dimensions: 1) For spatial resolution, the use of spectromicroscopic techniques with thin membranes will be extended and the development of near ambient pressure[182] or transmission[183] photoelectron microscopes can be envisioned to help overcome limitations from the challenging sample design and robustness. 2) For the time dimension, synchrotron-based pump-probe experiments[184] and ultra-fast laser setups[185,186] can yield picosecond resolution for the study of excited/intermediate states during charge generation and transfer. Alternatively, so-called "time-multiplexed" techniques may help overcome signal variations from spatial drift, changes in the background absorption, or incoming X-ray intensity to isolate weak signals during longer integration times.[187] For the complete understanding of solid-liquid interfaces, we envision that a combined approach of diverse operando probes is necessary, for example by combining XAS and XPS,[119] X-ray scattering and (standing wave) XPS,[188] or by combining X-ray based techniques with optical spectroscopies or scanning probe microscopies.

# Acknowledgement

This project has received funding from the European Research Council (ERC) under the European Union's Horizon 2020 research and innovation programme under grant agreement No. 101040669 – Interfaces at Work. This project was supported by the Royal Academy of Engineering under the Research Fellowship programme. The authors thank Qijun Che and Frank de Groot for joint XAS synchrotron activities and discussion.

# References

(1) Helmholtz, H. Studien Über Electrische Grenzschichten. *Annalen der Physik und Chemie* **1879**, *243* (7), 337–382. https://doi.org/10.1002/andp.18792430702.

(2) Daniell, J. F.; Miller, W. A. I. Additional Researches on the Electrolysis of Secondary Compounds. *Philos Trans R Soc Lond* **1844**, *134*, 1–19. https://doi.org/10.1098/rstl.1844.0001.

(3) Jerkiewicz, G. From Electrochemistry to Molecular-Level Research on the Solid—Liquid Electrochemical Interface. In *Solid-Liquid Electrochemical Interfaces*; Jerkiewicz, G., Soriaga, M. P., Wieckowski, A., Uosaki, K., Eds.; American Chemical Society: Washington, DC, 1997; Vol. 656, pp 1–12. https://doi.org/10.1021/bk-1997-0656.ch001.

(4) Björneholm, O.; Hansen, M. H.; Hodgson, A.; Liu, L.-M.; Limmer, D. T.; Michaelides, A.; Pedevilla, P.; Rossmeisl, J.; Shen, H.; Tocci, G.; Tyrode, E.; Walz, M.-M.; Werner, J.; Bluhm, H. Water at Interfaces. *Chem Rev* **2016**, *116* (13), 7698–7726. https://doi.org/10.1021/acs.chemrev.6b00045.

(5) Favaro, M.; Jeong, B.; Ross, P. N.; Yano, J.; Hussain, Z.; Liu, Z.; Crumlin, E. J. Unravelling the Electrochemical Double Layer by Direct Probing of the Solid/Liquid Interface. *Nat Commun* **2016**, *7* (May), 1–8. https://doi.org/10.1038/ncomms12695.

(6) Velasco-Velez, J. J.; Wu, C. H.; Pascal, T. A.; Wan, L. F.; Guo, J.; Prendergast, D.; Salmeron, M. The Structure of Interfacial Water on Gold Electrodes Studied by X-Ray Absorption Spectroscopy. *Science (1979)* **2014**, *346* (6211), 831–834. https://doi.org/10.1126/SCIENCE.1259437/SUPPL_FILE/VELASCO-VELEZ-SM.PDF.

(7) Wu, C. H.; Weatherup, R. S.; Salmeron, M. B. Probing Electrode/Electrolyte Interfaces in Situ by X-Ray Spectroscopies: Old Methods, New Tricks. *Physical Chemistry Chemical Physics* **2015**, *17* (45), 30229–30239. https://doi.org/10.1039/C5CP04058B.

(8) Toney, M. F.; Howard, J. N.; Richer, J.; Borges, G. L.; Gordon, J. G.; Melroy, O. R.; Wiesler, D. G.; Yee, D.; Sorensen, L. B. Voltage-Dependent Ordering of Water Molecules at an Electrode-Electrolyte Interface. *Nature* **1994**, *368*, 444–446.

(9) Weckhuysen, B. M. Snapshots of a Working Catalyst: Possibilities and Limitations of in Situ Spectroscopy in the Field of Heterogeneous Catalysis. *Chemical Communications* **2002**, No. 2, 97–110. https://doi.org/10.1039/b107686h.

(10) Bañares, M. A.; Wachs, I. E. Molecular Structures of Supported Metal Oxide Catalysts under Different Environments. *Journal of Raman Spectroscopy* **2002**, *33* (5), 359–380. https://doi.org/10.1002/jrs.866.

(11) Shin, H.; Yoo, J. M.; Sung, Y.; Chung, D. Y. Dynamic Electrochemical Interfaces for Energy Conversion and Storage. *JACS Au* **2022**. https://doi.org/10.1021/jacsau.2c00385.

(12) Kibsgaard, J.; Chorkendorff, I. Considerations for the Scaling-up of Water Splitting Catalysts. *Nat Energy* **2019**, *4* (6), 430–433. https://doi.org/10.1038/s41560-019-0407-1.

(13) Zwaschka, G.; Nahalka, I.; Marchioro, A.; Tong, Y.; Roke, S.; Campen, R. K. Imaging the Heterogeneity of the Oxygen Evolution Reaction on Gold Electrodes Operando: Activity Is Highly Local. *ACS Catal* **2020**, *10* (11), 6084–6093. https://doi.org/10.1021/acscatal.0c01177.

(14) Baeumer, C.; Li, J.; Lu, Q.; Liang, A. Y.-L.; Jin, L.; Martins, H. P.; Duchoň, T.; Glöß, M.; Gericke, S. M.; Wohlgemuth, M. A.; Giesen, M.; Penn, E. E.; Dittmann, R.; Gunkel, F.; Waser, R.; Bajdich, M.; Nemšák, S.; Mefford, J. T.; Chueh, W. C. Tuning Electrochemically Driven Surface Transformation in Atomically Flat LaNiO3 Thin Films for Enhanced Water Electrolysis. *Nat Mater* **2021**, *accepted*. https://doi.org/10.1038/s41563-020-00877-1.

(15) Chung, D. Y.; Lopes, P. P.; Farinazzo Bergamo Dias Martins, P.; He, H.; Kawaguchi, T.; Zapol, P.; You, H.; Tripkovic, D.; Strmcnik, D.; Zhu, Y.; Seifert, S.; Lee, S.; Stamenkovic, V. R.; Markovic, N. M. Dynamic Stability of Active Sites in Hydr(Oxy)Oxides for the Oxygen Evolution Reaction. *Nat Energy* **2020**, *5* (3), 222–230. https://doi.org/10.1038/s41560-020-0576-y.

(16) Salmeron, M. From Surfaces to Interfaces: Ambient Pressure XPS and Beyond. *Top Catal* **2018**, *61* (20), 2044–2051. https://doi.org/10.1007/s11244-018-1069-0.

(17) Starr, D. E.; Liu, Z.; Hävecker, M.; Knop-Gericke, A.; Bluhm, H. Investigation of Solid/Vapor Interfaces Using Ambient Pressure X-Ray Photoelectron Spectroscopy. *Chem Soc Rev* **2013**, *42* (13), 5833. https://doi.org/10.1039/c3cs60057b.

(18) Zhu, K.; Zhu, X.; Yang, W. Application of In Situ Techniques for the Characterization of NiFe-Based Oxygen Evolution Reaction (OER) Electrocatalysts. *Angewandte Chemie International Edition* **2019**, *58* (5), 1252–1265. https://doi.org/10.1002/anie.201802923.

(19) Choi, Y.-W.; Mistry, H.; Roldan Cuenya, B. New Insights into Working Nanostructured Electrocatalysts through Operando Spectroscopy and Microscopy. *Curr Opin Electrochem* **2017**, *1* (1), 95–103. https://doi.org/10.1016/j.coelec.2017.01.004.

(20) Traulsen, M. L.; Chatzichristodoulou, C.; Hansen, K. v.; Kuhn, L. T.; Holtappels, P.; Mogensen, M. B. Need for In Operando Characterization of Electrochemical Interface Features. *ECS Trans* **2015**, *66* (2), 3–20. https://doi.org/10.1149/06602.0003ecst.

(21) Itkis, D. M.; Velasco-Velez, J. J.; Knop-Gericke, A.; Vyalikh, A.; Avdeev, M. v.; Yashina, L. v. Probing Operating Electrochemical Interfaces by Photons and Neutrons. *ChemElectroChem* **2015**, *2* (10), 1427–1445. https://doi.org/10.1002/celc.201500155.

(22) Liu, D.; Shadike, Z.; Lin, R.; Qian, K.; Li, H.; Li, K.; Wang, S.; Yu, Q.; Liu, M.; Ganapathy, S.; Qin, X.; Yang, Q.; Wagemaker, M.; Kang, F.; Yang, X.; Li, B. Review of Recent Development of In Situ/Operando Characterization Techniques for Lithium Battery Research. *Advanced Materials* **2019**, *31* (28), 1806620. https://doi.org/10.1002/adma.201806620.

(23) Mul, G.; de Groot, F.; Mojet-Mol, B.; Tromp, M. Characterization of Catalysts. In *Catalysis: An Integrated Textbook for Students*; Hanefeld, U., Lefferts, L., Eds.; Wiley-VCH: Weinheim, Germany, 2017; pp 271–314.

(24) Kolmakov, A.; Gregoratti, L.; Kiskinova, M.; Günther, S. Recent Approaches for Bridging the Pressure Gap in Photoelectron Microspectroscopy. *Top Catal* **2016**, *59* (5–7), 448–468. https://doi.org/10.1007/s11244-015-0519-1.

(25) Axnanda, S.; Crumlin, E. J.; Mao, B.; Rani, S.; Chang, R.; Karlsson, P. G.; Edwards, M. O. M.; Lundqvist, M.; Moberg, R.; Ross, P.; Hussain, Z.; Liu, Z. Using "Tender" X-Ray Ambient Pressure X-Ray Photoelectron Spectroscopy as A Direct Probe of Solid-Liquid Interface. *Sci Rep* **2015**, *5* (1), 9788. https://doi.org/10.1038/srep09788.

(26) Favaro, M.; Abdi, F.; Crumlin, E.; Liu, Z.; van de Krol, R.; Starr, D. Interface Science Using Ambient Pressure Hard X-Ray Photoelectron Spectroscopy. *Surfaces* **2019**, *2* (1), 78–99. https://doi.org/10.3390/surfaces2010008.

(27) Carbonio, E. A.; Velasco-Velez, J.-J.; Schlögl, R.; Knop-Gericke, A. Perspective—Outlook on Operando Photoelectron and Absorption Spectroscopy to Probe Catalysts at the Solid-Liquid Electrochemical Interface. *J Electrochem Soc* **2020**, *167* (5), 054509. https://doi.org/10.1149/1945-7111/ab68d2.

(28) Stoerzinger, K. A.; Favaro, M.; Ross, P. N.; Hussain, Z.; Liu, Z.; Yano, J.; Crumlin, E. J. Stabilizing the Meniscus for Operando Characterization of Platinum During the Electrolyte-Consuming Alkaline Oxygen Evolution Reaction. *Top Catal* **2018**, *61* (20), 2152–2160. https://doi.org/10.1007/s11244-018-1063-6.

(29) Novotny, Z.; Aegerter, D.; Comini, N.; Tobler, B.; Artiglia, L.; Maier, U.; Moehl, T.; Fabbri, E.; Huthwelker, T.; Schmidt, T. J.; Ammann, M.; van Bokhoven, J. A.; Raabe, J.; Osterwalder, J. Probing the Solid–Liquid Interface with Tender x Rays: A New Ambient-Pressure x-Ray Photoelectron Spectroscopy Endstation at the Swiss Light Source. *Review of Scientific Instruments* **2020**, *91* (2), 023103. https://doi.org/10.1063/1.5128600.

(30) Karslıoğlu, O.; Nemšák, S.; Zegkinoglou, I.; Shavorskiy, A.; Hartl, M.; Salmassi, F.; Gullikson, E. M.; Ng, M. L.; Rameshan, Ch.; Rude, B.; Bianculli, D.; Cordones, A. A.; Axnanda, S.; Crumlin, E. J.; Ross, P. N.; Schneider, C. M.; Hussain, Z.; Liu, Z.; Fadley, C. S.; Bluhm, H. Aqueous Solution/Metal Interfaces Investigated in Operando by Photoelectron Spectroscopy. *Faraday Discuss* **2015**, *180* (0), 35–53. https://doi.org/10.1039/C5FD00003C.

(31) Nemšák, S.; Shavorskiy, A.; Karslioglu, O.; Zegkinoglou, I.; Rattanachata, A.; Conlon, C. S.; Keqi, A.; Greene, P. K.; Burks, E. C.; Salmassi, F.; Gullikson, E. M.; Yang, S.-H.; Liu, K.; Bluhm, H.; Fadley, C. S. Concentration and Chemical-State Profiles at Heterogeneous Interfaces with Sub-Nm Accuracy from Standing-Wave Ambient-Pressure Photoemission. *Nat Commun* **2014**, *5* (1), 5441. https://doi.org/10.1038/ncomms6441.

(32) Gray, A. X.; Nemšák, S.; Fadley, C. S. Combining Hard and Soft X-Ray Photoemission with Standing-Wave Excitation, Resonant Excitation, and Angular Resolution. *Synchrotron Radiat News* **2018**, *31* (4), 42–49. https://doi.org/10.1080/08940886.2018.1483659.

(33) Nemšák, S.; Gray, A. X.; Fadley, C. S. Standing-Wave and Resonant Soft- and Hard-X-Ray Photoelectron Spectroscopy of Oxide Interfaces. In *Spectroscopy of Complex Oxide Interfaces: Photoemission and Related Spectroscopies*; Cancellieri, C., Strocov, V. N., Eds.; Springer International Publishing: Cham, 2018; pp 153–179. https://doi.org/10.1007/978-3-319-74989-1_7.

(34) Timoshenko, J.; Roldan Cuenya, B. In Situ/ Operando Electrocatalyst Characterization by X-Ray Absorption Spectroscopy. *Chem Rev* **2021**, *121* (2), 882–961. https://doi.org/10.1021/ACS.CHEMREV.0C00396/ASSET/IMAGES/MEDIUM/CR0C00396_0029.GIF.

(35) Bak, S.-M.; Shadike, Z.; Lin, R.; Yu, X.; Yang, X.-Q. In Situ/Operando Synchrotron-Based X-Ray Techniques for Lithium-Ion Battery Research. *NPG Asia Mater* **2018**, *10* (7), 563–580. https://doi.org/10.1038/s41427-018-0056-z.

(36) Gorlin, Y.; Lassalle-Kaiser, B.; Benck, J. D.; Gul, S.; Webb, S. M.; Yachandra, V. K.; Yano, J.; Jaramillo, T. F. In Situ X-Ray Absorption Spectroscopy Investigation of a Bifunctional Manganese Oxide Catalyst with High Activity for Electrochemical Water Oxidation and Oxygen Reduction. *J Am Chem Soc* **2013**, *135* (23), 8525–8534. https://doi.org/10.1021/ja3104632.

(37) Starr, D. E.; Favaro, M.; Abdi, F. F.; Bluhm, H.; Crumlin, E. J.; van de Krol, R. Combined Soft and Hard X-Ray Ambient Pressure Photoelectron Spectroscopy Studies of Semiconductor/Electrolyte Interfaces. *J Electron Spectros Relat Phenomena* **2017**, *221*, 106–115. https://doi.org/10.1016/j.elspec.2017.05.003.

(38) Sushko, P. v.; Chambers, S. A. Extracting Band Edge Profiles at Semiconductor Heterostructures from Hard-x-Ray Core-Level Photoelectron Spectra. *Sci Rep* **2020**, *10* (1), 13028. https://doi.org/10.1038/s41598-020-69658-9.

(39) Müller, M.; Nemšák, S.; Plucinski, L.; Schneider, C. M. Functional Materials for Information and Energy Technology: Insights by Photoelectron Spectroscopy. *J Electron Spectros Relat Phenomena* **2016**, *208*, 24–32. https://doi.org/10.1016/j.elspec.2015.08.003.

(40) Fadley, C. S. Hard X-Ray Photoemission: An Overview and Future Perspective. In *Hard X-ray Photoelectron Spectroscopy ( HAXPES )*; Woicik, J. C., Ed.; Heidelberg, New York, 2016; pp 1–34.

(41) Busse, P.; Yin, Z.; Mierwaldt, D.; Scholz, J.; Kressdorf, B.; Glaser, L.; Miedema, P. S.; Rothkirch, A.; Viefhaus, J.; Jooss, C.; Techert, S.; Risch, M. Probing the Surface of La 0.6 Sr 0.4 MnO 3 in Water Vapor by In Situ Photon-In/Photon-Out Spectroscopy. *The Journal of Physical Chemistry C* **2020**, *124* (14), 7893–7902. https://doi.org/10.1021/acs.jpcc.0c00840.

(42) Fadley, C. S. X-Ray Photoelectron Spectroscopy: Progress and Perspectives. *J Electron Spectros Relat Phenomena* **2010**, *178–179*, 2–32. https://doi.org/10.1016/j.elspec.2010.01.006.

(43) Briggs, D.; Grant, J. T. *Surface Analysis by Auger and X-Ray Photoelectron Spectroscopy*; IM Publications: Chichester, West Sussex, U.K., 2003.

(44) Thapa, S.; Paudel, R.; Blanchet, M. D.; Gemperline, P. T.; Comes, R. B. Probing Surfaces and Interfaces in Complex Oxide Films via in Situ X-Ray Photoelectron Spectroscopy. *J Mater Res* **2020**, 1–26. https://doi.org/10.1557/jmr.2020.261.

(45) Baer, D. R.; Artyushkova, K.; Richard Brundle, C.; Castle, J. E.; Engelhard, M. H.; Gaskell, K. J.; Grant, J. T.; Haasch, R. T.; Linford, M. R.; Powell, C. J.; Shard, A. G.; Sherwood, P. M. A.; Smentkowski, V. S. Practical Guides for X-Ray Photoelectron Spectroscopy: First Steps in Planning, Conducting, and Reporting XPS Measurements. *Journal of Vacuum Science & Technology A* **2019**, *37* (3), 031401. https://doi.org/10.1116/1.5065501.

(46) Powell, C. J. Practical Guide for Inelastic Mean Free Paths, Effective Attenuation Lengths, Mean Escape Depths, and Information Depths in x-Ray Photoelectron Spectroscopy. *Journal of Vacuum Science & Technology A* **2020**, *38* (2), 023209. https://doi.org/10.1116/1.5141079.

(47) Chambers, S. A.; Wang, L.; Baer, D. R. Introductory Guide to the Application of XPS to Epitaxial Films and Heterostructures. *Journal of Vacuum Science & Technology A* **2020**, *38* (6), 061201. https://doi.org/10.1116/6.0000465.

(48) Tougaard, S. Practical Guide to the Use of Backgrounds in Quantitative XPS. *Journal of Vacuum Science & Technology A* **2021**, *39* (1), 011201. https://doi.org/10.1116/6.0000661.

(49) Fleischmann, M.; Mao, B. W. In-Situ X-Ray Diffraction Studies of Pt Electrode/Solution Interfaces. *J Electroanal Chem Interfacial Electrochem* **1987**, *229* (1–2), 125–139. https://doi.org/10.1016/0022-0728(87)85135-5.

(50) Reikowski, F.; Maroun, F.; Pacheco, I.; Wiegmann, T.; Allongue, P.; Stettner, J.; Magnussen, O. M. Operando Surface X-Ray Diffraction Studies of Structurally Defined $Co_3O_4$ and CoOOH Thin Films during Oxygen Evolution. *ACS Catal* **2019**, *9* (5), 3811–3821. https://doi.org/10.1021/acscatal.8b04823.

(51) Drnec, J.; Ruge, M.; Reikowski, F.; Rahn, B.; Carlà, F.; Felici, R.; Stettner, J.; Magnussen, O. M.; Harrington, D. A. Initial Stages of Pt(111) Electrooxidation: Dynamic and Structural Studies by Surface X-Ray Diffraction. *Electrochim Acta* **2017**, *224*. https://doi.org/10.1016/j.electacta.2016.12.028.

(52) Magnussen, O. M.; Krug, K.; Ayyad, A. H.; Stettner, J. In Situ Diffraction Studies of Electrode Surface Structure during Gold Electrodeposition. *Electrochim Acta* **2008**, *53* (9), 3449–3458. https://doi.org/10.1016/j.electacta.2007.10.037.

(53) Rao, R. R.; Kolb, M. J.; Giordano, L.; Pedersen, A. F.; Katayama, Y.; Hwang, J.; Mehta, A.; You, H.; Lunger, J. R.; Zhou, H.; Halck, N. B.; Vegge, T.; Chorkendorff, I.; Stephens, I. E. L.; Shao-Horn, Y. Operando Identification of Site-Dependent Water Oxidation Activity on Ruthenium Dioxide Single-Crystal Surfaces. *Nat Catal* **2020**, *3* (6), 516–525. https://doi.org/10.1038/s41929-020-0457-6.

(54) Rao, R. R.; Kolb, M. J.; Halck, N. B.; Pedersen, A. F.; Mehta, A.; You, H.; Stoerzinger, K. A.; Feng, Z.; Hansen, H. A.; Zhou, H.; Giordano, L.; Rossmeisl, J.; Vegge, T.; Chorkendorff, I.; Stephens, I. E. L.; Shao-Horn, Y. Towards Identifying the Active Sites on RuO2 (110) in Catalyzing Oxygen Evolution. *Energy Environ Sci* **2017**, *10* (12), 2626–2637. https://doi.org/10.1039/C7EE02307C.

(55) Disa, A. S.; Walker, F. J.; Ahn, C. H. High-Resolution Crystal Truncation Rod Scattering: Application to Ultrathin Layers and Buried Interfaces. *Adv Mater Interfaces* **2020**, *7* (6), 1901772. https://doi.org/10.1002/admi.201901772.

(56) de Groot, F. High-Resolution X-Ray Emission and X-Ray Absorption Spectroscopy. *Chem Rev* **2001**, *101* (6), 1779–1808. https://doi.org/10.1021/cr9900681.

(57) Stöhr, J. *NEXAFS Spectroscopy*; Springer Series in Surface Sciences; Springer Berlin Heidelberg: Berlin, Heidelberg, 1992; Vol. 25. https://doi.org/10.1007/978-3-662-02853-7.

(58) Shulman, G. R.; Yafet, Y.; Eisenberger, P.; Blumberg, W. E. Observations and Interpretation of X-Ray Absorption Edges in Iron Compounds and Proteins. *Proceedings of the National Academy of Sciences* **1976**, *73* (5), 1384–1388. https://doi.org/10.1073/pnas.73.5.1384.

(59) Yano, J.; Yachandra, V. K. X-Ray Absorption Spectroscopy. *Photosynth Res* **2009**, *102* (2–3), 241–254. https://doi.org/10.1007/s11120-009-9473-8.

(60) Johnston, R. W.; Tomboulian, D. H. Absorption Spectrum of Beryllium in the Neighborhood of the $K$ Edge. *Physical Review* **1954**, *94* (6), 1585–1589. https://doi.org/10.1103/PhysRev.94.1585.

(61) Henke, B. L.; Gullikson, E. M.; Davis, J. C. X-Ray Interactions: Photoabsorption, Scattering, Transmission, and Reflection at E = 50-30,000 EV, Z = 1-92. *At Data Nucl Data Tables* **1993**, *54* (2), 181–342. https://doi.org/10.1006/adnd.1993.1013.

(62) Friebel, D.; Miller, D. J.; Ogasawara, H.; Anniyev, T.; Bergmann, U.; Bargar, J.; Nilsson, A. In Situ GIXAFS and HERFD-XAS Studies of a Pt-Modified Rh ( 111 ) Electrode. **2009**, No. 111.

(63) Drevon, D.; Görlin, M.; Chernev, P.; Xi, L.; Dau, H.; Lange, K. M. Uncovering The Role of Oxygen in Ni-Fe(OxHy) Electrocatalysts Using In Situ Soft X-Ray Absorption Spectroscopy during the Oxygen Evolution Reaction. *Sci Rep* **2019**, *9* (1), 1–11. https://doi.org/10.1038/s41598-018-37307-x.

(64) Velasco-Vélez, J.-J.; Falling, L. J.; Bernsmeier, D.; Sear, M. J.; Clark, P. C. J.; Chan, T.-S.; Stotz, E.; Hävecker, M.; Kraehnert, R.; Knop-Gericke, A.; Chuang, C.-H.; Starr, D. E.; Favaro, M.; Mom, R. v. A Comparative Study of Electrochemical Cells for in Situ X-Ray Spectroscopies in the Soft and Tender x-Ray Range. *J Phys D Appl Phys* **2021**, *54* (12), 124003. https://doi.org/10.1088/1361-6463/abd2ed.

(65) Velasco-Velez, J. J.; Wu, C. H.; Wang, B. Y.; Sun, Y.; Zhang, Y.; Guo, J.-H.; Salmeron, M. Polarized X-Ray Absorption Spectroscopy Observation of Electronic and Structural Changes of Chemical Vapor Deposition Graphene in Contact with Water. *The Journal of Physical Chemistry C* **2014**, *118* (44), 25456–25459. https://doi.org/10.1021/jp507405z.

(66) Kolmakov, A.; Dikin, D. A.; Cote, L. J.; Huang, J.; Abyaneh, M. K.; Amati, M.; Gregoratti, L.; Günther, S.; Kiskinova, M. Graphene Oxide Windows for in Situ Environmental Cell Photoelectron Spectroscopy. *Nat Nanotechnol* **2011**, *6* (10), 651–657. https://doi.org/10.1038/nnano.2011.130.

(67) Kraus, J.; Reichelt, R.; Günther, S.; Gregoratti, L.; Amati, M.; Kiskinova, M.; Yulaev, A.; Vlassiouk, I.; Kolmakov, A. Photoelectron Spectroscopy of Wet and Gaseous Samples through Graphene Membranes. *Nanoscale* **2014**, *6* (23), 14394–14403. https://doi.org/10.1039/C4NR03561E.

(68) Lu, Y.-H.; Morales, C.; Zhao, X.; van Spronsen, M. A.; Baskin, A.; Prendergast, D.; Yang, P.; Bechtel, H. A.; Barnard, E. S.; Ogletree, D. F.; Altoe, V.; Soriano, L.; Schwartzberg, A. M.; Salmeron, M. Ultrathin Free-Standing Oxide Membranes for Electron and Photon Spectroscopy Studies of Solid–Gas and Solid–Liquid Interfaces. *Nano Lett* **2020**, *20* (9), 6364–6371. https://doi.org/10.1021/acs.nanolett.0c01801.

(69) Falling, L. J.; Mom, R. v.; Sandoval Diaz, L. E.; Nakhaie, S.; Stotz, E.; Ivanov, D.; Hävecker, M.; Lunkenbein, T.; Knop-Gericke, A.; Schlögl, R.; Velasco-Vélez, J.-J. Graphene-Capped Liquid Thin Films for Electrochemical Operando X-Ray Spectroscopy and Scanning Electron Microscopy. *ACS Appl Mater Interfaces* **2020**, *12* (33), 37680–37692. https://doi.org/10.1021/acsami.0c08379.

(70) Liu, J.; Jia, E.; Stoerzinger, K. A.; Wang, L.; Wang, Y.; Yang, Z.; Shen, D.; Engelhard, M. H.; Bowden, M. E.; Zhu, Z.; Chambers, S. A.; Du, Y. Dynamic Lattice Oxygen Participation on Perovskite LaNiO 3 during Oxygen Evolution Reaction. *The Journal of Physical Chemistry C* **2020**, *124* (28), 15386–15390. https://doi.org/10.1021/acs.jpcc.0c04808.

(71) Pfeifer, V.; Jones, T. E.; Velasco Vélez, J. J.; Arrigo, R.; Piccinin, S.; Hävecker, M.; Knop-Gericke, A.; Schlögl, R. In Situ Observation of Reactive Oxygen Species Forming on Oxygen-Evolving Iridium Surfaces. *Chem Sci* **2017**, *8* (3), 2143–2149. https://doi.org/10.1039/C6SC04622C.

(72)  Risch, M.; Morales, D. M.; Villalobos, J.; Antipin, D. What X-Ray Absorption Spectroscopy Can Tell Us About the Active State of Earth-Abundant Electrocatalysts for the Oxygen Evolution Reaction. *Angewandte Chemie International Edition* **2022**. https://doi.org/10.1002/anie.202211949.

(73)  Liu, Y.-S.; Feng, X.; Glans, P.-A.; Guo, J. In-Situ/Operando Soft x-Ray Spectroscopy Characterization of Energy and Catalytic Materials. *Solar Energy Materials and Solar Cells* **2020**, *208* (2), 110432. https://doi.org/10.1016/j.solmat.2020.110432.

(74)  Friebel, D.; Louie, M. W.; Bajdich, M.; Sanwald, K. E.; Cai, Y.; Wise, A. M.; Cheng, M. J.; Sokaras, D.; Weng, T. C.; Alonso-Mori, R.; Davis, R. C.; Bargar, J. R.; Nørskov, J. K.; Nilsson, A.; Bell, A. T. Identification of Highly Active Fe Sites in (Ni,Fe)OOH for Electrocatalytic Water Splitting. *J Am Chem Soc* **2015**, *137* (3), 1305–1313. https://doi.org/10.1021/JA511559D/SUPPL_FILE/JA511559D_SI_001.PDF.

(75)  Bates, M. K.; Jia, Q.; Doan, H.; Liang, W.; Mukerjee, S. Charge-Transfer Effects in Ni-Fe and Ni-Fe-Co Mixed-Metal Oxides for the Alkaline Oxygen Evolution Reaction. *ACS Catal* **2016**, *6* (1), 155–161. https://doi.org/10.1021/ACSCATAL.5B01481/ASSET/IMAGES/LARGE/CS-2015-014812_0006.JPEG.

(76)  Ismail, A. S. M.; Garcia-Torregrosa, I.; Vollenbroek, J. C.; Folkertsma, L.; Bomer, J. G.; Haarman, T.; Ghiasi, M.; Schellhorn, M.; Nachtegaal, M.; Odijk, M.; van den Berg, A.; Weckhuysen, B. M.; de Groot, F. M. F. Detection of Spontaneous FeOOH Formation at the Hematite/Ni(Fe)OOH Interface during Photoelectrochemical Water Splitting by Operando X-Ray Absorption Spectroscopy. *ACS Catal* **2021**, *11* (19), 12324–12335. https://doi.org/10.1021/ACSCATAL.1C02566/ASSET/IMAGES/LARGE/CS1C02566_0006.JPEG.

(77)  Wang, D.; Zhou, J.; Hu, Y.; Yang, J.; Han, N.; Li, Y.; Sham, T. K. In Situ X-Ray Absorption Near-Edge Structure Study of Advanced NiFe(OH)x Electrocatalyst on Carbon Paper for Water Oxidation. *Journal of Physical Chemistry C* **2015**, *119* (34), 19573–19583. https://doi.org/10.1021/ACS.JPCC.5B02685/SUPPL_FILE/JP5B02685_SI_001.PDF.

(78)  van Spronsen, M. A.; Zhao, X.; Jaugstetter, M.; Escudero, C.; Duchoň, T.; Hunt, A.; Waluyo, I.; Yang, P.; Tschulik, K.; Salmeron, M. B. Interface Sensitivity in Electron/Ion Yield X-Ray Absorption Spectroscopy: The TiO2-H2O Interface. *Journal of Physical Chemistry Letters* **2021**, *12* (41), 10212–10217. https://doi.org/10.1021/ACS.JPCLETT.1C02115/ASSET/IMAGES/LARGE/JZ1C02115_0004.JPEG.

(79)  Schön, D.; Xiao, J.; Golnak, R.; Tesch, M. F.; Winter, B.; Velasco-Velez, J.-J.; Aziz, E. F. Introducing Ionic-Current Detection for X-Ray Absorption Spectroscopy in Liquid Cells. *J Phys Chem Lett* **2017**, *8* (9), 2087–2092. https://doi.org/10.1021/acs.jpclett.7b00646.

(80)  Streibel, V.; Velasco-Vélez, J. J.; Teschner, D.; Carbonio, E. A.; Knop-Gericke, A.; Schlögl, R.; Jones, T. E. Merging Operando and Computational X-Ray Spectroscopies to Study the Oxygen Evolution Reaction. *Curr Opin Electrochem* **2022**, *35*, 101039. https://doi.org/10.1016/j.coelec.2022.101039.

(81)  Nørskov, J. K.; Rossmeisl, J.; Logadottir, A.; Lindqvist, L.; Kitchin, J. R.; Bligaard, T.; Jónsson, H. Origin of the Overpotential for Oxygen Reduction at a Fuel-Cell Cathode. *J Phys Chem B* **2004**, *108* (46), 17886–17892. https://doi.org/10.1021/jp047349j.

(82)  Haverkort, M. W. Quanty for Core Level Spectroscopy - Excitons, Resonances and Band Excitations in Time and Frequency Domain. *J Phys Conf Ser* **2016**, *712* (1), 012001. https://doi.org/10.1088/1742-6596/712/1/012001.

(83)  Stavitski, E.; de Groot, F. M. F. The CTM4XAS Program for EELS and XAS Spectral Shape Analysis of Transition Metal L Edges. *Micron* **2010**, *41* (7), 687–694. https://doi.org/10.1016/J.MICRON.2010.06.005.

(84)  Rehr, J. J.; Albers, R. C. Theoretical Approaches to X-Ray Absorption Fine Structure. *Rev Mod Phys* **2000**, *72* (3), 621. https://doi.org/10.1103/RevModPhys.72.621.

(85)  Rehr, J. J.; Kas, J. J.; Prange, M. P.; Sorini, A. P.; Takimoto, Y.; Vila, F. Ab Initio Theory and Calculations of X-Ray Spectra. *C R Phys* **2009**, *10* (6), 548–559. https://doi.org/10.1016/J.CRHY.2008.08.004.

(86)  Werner, W. S. M. Questioning a Universal Law for Electron Attenuation. *Physics (College Park Md)* **2019**, *12*, 93. https://doi.org/10.1103/Physics.12.93.

(87)  Ottosson, N.; Faubel, M.; Bradforth, S. E.; Jungwirth, P.; Winter, B. Photoelectron Spectroscopy of Liquid Water and Aqueous Solution: Electron Effective Attenuation Lengths and Emission-Angle Anisotropy. *J Electron Spectros Relat Phenomena* **2010**, *177* (2–3), 60–70. https://doi.org/10.1016/j.elspec.2009.08.007.

(88)  Tanuma, S.; Powell, C. J.; Penn, D. R. Calculations of Electron Inelastic Mean Free Paths. V. Data for 14 Organic Compounds over the 50-2000 EV Range. *Surface and Interface Analysis* **1994**, *21* (3), 165–176. https://doi.org/10.1002/sia.740210302.

(89)  Siegbahn, H.; Siegbahn, K. ESCA Applied to Liquids. *J Electron Spectros Relat Phenomena* **1973**, *2* (3), 319–325. https://doi.org/10.1016/0368-2048(73)80023-4.

(90) Siegbahn, H.; Svensson, S.; Lundholm, M. A New Method for ESCA Studies of Liquid-Phase Samples. *J Electron Spectros Relat Phenomena* **1981**, *24* (2), 205–213. https://doi.org/10.1016/0368-2048(81)80007-2.

(91) Ogletree, D. F.; Bluhm, H.; Lebedev, G.; Fadley, C. S.; Hussain, Z.; Salmeron, M. A Differentially Pumped Electrostatic Lens System for Photoemission Studies in the Millibar Range. *Review of Scientific Instruments* **2002**, *73* (11), 3872–3877. https://doi.org/10.1063/1.1512336.

(92) Bluhm, H.; Hävecker, M.; Knop-Gericke, A.; Kleimenov, E.; Schlögl, R.; Teschner, D.; Bukhtiyarov, V. I.; Ogletree, D. F.; Salmeron, M. Methanol Oxidation on a Copper Catalyst Investigated Using in Situ X-Ray Photoelectron Spectroscopy †. *J Phys Chem B* **2004**, *108* (38), 14340–14347. https://doi.org/10.1021/jp040080j.

(93) Bluhm, H. Preface to the Special Issue of Topics in Catalysis on Ambient Pressure X-Ray Photoelectron Spectroscopy. *Top Catal* **2016**, *59* (5–7), 403–404. https://doi.org/10.1007/s11244-015-0514-6.

(94) Shavorskiy, A.; Karslioglu, O.; Zegkinoglou, I.; Bluhm, H. Synchrotron-Based Ambient Pressure X-Ray Photoelectron Spectroscopy. *Synchrotron Radiat News* **2014**, *27* (2), 14–23. https://doi.org/10.1080/08940886.2014.889547.

(95) Newberg, J. T.; Åhlund, J.; Arble, C.; Goodwin, C.; Khalifa, Y.; Broderick, A. A Lab-Based Ambient Pressure x-Ray Photoelectron Spectrometer with Exchangeable Analysis Chambers. *Review of Scientific Instruments* **2015**, *86* (8), 085113. https://doi.org/10.1063/1.4928498.

(96) Arble, C.; Jia, M.; Newberg, J. T. Lab-Based Ambient Pressure X-Ray Photoelectron Spectroscopy from Past to Present. *Surf Sci Rep* **2018**, *73* (2), 37–57. https://doi.org/10.1016/j.surfrep.2018.02.002.

(97) Roy, K.; Artiglia, L.; van Bokhoven, J. A. Ambient Pressure Photoelectron Spectroscopy: Opportunities in Catalysis from Solids to Liquids and Introducing Time Resolution. *ChemCatChem* **2018**, *10* (4), 666–682. https://doi.org/10.1002/cctc.201701522.

(98) Stoerzinger, K. A.; Hong, W. T.; Crumlin, E. J.; Bluhm, H.; Shao-Horn, Y. Insights into Electrochemical Reactions from Ambient Pressure Photoelectron Spectroscopy. *Acc Chem Res* **2015**, *48* (11), 2976–2983. https://doi.org/10.1021/acs.accounts.5b00275.

(99) Schnadt, J.; Knudsen, J.; Johansson, N. Present and New Frontiers in Materials Research by Ambient Pressure X-Ray Photoelectron Spectroscopy. *Journal of Physics: Condensed Matter* **2020**, *32* (41), 413003. https://doi.org/10.1088/1361-648X/ab9565.

(100) Cai, J.; Dong, Q.; Han, Y.; Mao, B.-H.; Zhang, H.; Karlsson, P. G.; Åhlund, J.; Tai, R.-Z.; Yu, Y.; Liu, Z. An APXPS Endstation for Gas–Solid and Liquid–Solid Interface Studies at SSRF. *Nuclear Science and Techniques* **2019**, *30* (5), 81. https://doi.org/10.1007/s41365-019-0608-0.

(101) Wang, C.-H.; Liu, B.-H.; Yang, Y.-W. Ambient Pressure X-Ray Photoelectron Spectroscopy (APXPS): Present Status and Future Development at NSRRC. *Synchrotron Radiat News* **2022**, *35* (3), 48–53. https://doi.org/10.1080/08940886.2022.2082182.

(102) Zhu, S.; Scardamaglia, M.; Kundsen, J.; Sankari, R.; Tarawneh, H.; Temperton, R.; Pickworth, L.; Cavalca, F.; Wang, C.; Tissot, H.; Weissenrieder, J.; Hagman, B.; Gustafson, J.; Kaya, S.; Lindgren, F.; Källquist, I.; Maibach, J.; Hahlin, M.; Boix, V.; Gallo, T.; Rehman, F.; D'Acunto, G.; Schnadt, J.; Shavorskiy, A. HIPPIE: A New Platform for Ambient-Pressure X-Ray Photoelectron Spectroscopy at the MAX IV Laboratory. *J Synchrotron Radiat* **2021**, *28* (2), 624–636. https://doi.org/10.1107/S160057752100103X.

(103) Starr, D. E.; Hävecker, M.; Knop-Gericke, A.; Favaro, M.; Vadilonga, S.; Mertin, M.; Reichardt, G.; Schmidt, J.; Siewert, F.; Schulz, R.; Viefhaus, J.; Jung, C.; van de Krol, R. The Berlin Joint Lab for Electrochemical Interfaces, BElChem: A Facility for In-Situ and Operando NAP-XPS and NAP-HAXPES Studies of Electrochemical Interfaces at BESSY II. *Synchrotron Radiat News* **2022**, *35* (3), 54–60. https://doi.org/10.1080/08940886.2022.2082209.

(104) Bockris, J. O.; Cahan, B. D. Effect of a Finite-Contact-Angle Meniscus on Kinetics in Porous Electrode Systems. *J Chem Phys* **1969**, *50* (3), 1307–1324. https://doi.org/10.1063/1.1671193.

(105) Weingarth, D.; Foelske-Schmitz, A.; Wokaun, A.; Kötz, R. In Situ Electrochemical XPS Study of the Pt/[EMIM][BF4] System. *Electrochem commun* **2011**, *13* (6), 619–622. https://doi.org/10.1016/j.elecom.2011.03.027.

(106) Booth, S. G.; Tripathi, A. M.; Strashnov, I.; Dryfe, R. A. W.; Walton, A. S. The Offset Droplet: A New Methodology for Studying the Solid/Water Interface Using x-Ray Photoelectron Spectroscopy. *Journal of Physics: Condensed Matter* **2017**, *29* (45), 454001. https://doi.org/10.1088/1361-648X/aa8b92.

(107) Byrne, C.; Zahra, K. M.; Dhaliwal, S.; Grinter, D. C.; Roy, K.; Garzon, W. Q.; Held, G.; Thornton, G.; Walton, A. S. A Combined Laboratory and Synchrotron In-Situ Photoemission Study of the Rutile $TiO_2$ (110)/Water Interface. *J Phys D Appl Phys* **2021**, *54* (19), 194001. https://doi.org/10.1088/1361-6463/abddfb.

(108) Favaro, M.; Valero-Vidal, C.; Eichhorn, J.; Toma, F. M.; Ross, P. N.; Yano, J.; Liu, Z.; Crumlin, E. J. Elucidating the Alkaline Oxygen Evolution Reaction Mechanism on Platinum. *J Mater Chem A Mater* **2017**, *5* (23), 11634–11643. https://doi.org/10.1039/C7TA00409E.

(109) Stoerzinger, K. A.; Renshaw Wang, X.; Hwang, J.; Rao, R. R.; Hong, W. T.; Rouleau, C. M.; Lee, D.; Yu, Y.; Crumlin, E. J.; Shao-Horn, Y. Speciation and Electronic Structure of $La_{1-x}Sr_xCoO_{3-\delta}$ During Oxygen Electrolysis. *Top Catal* **2018**, *61* (20), 2161–2174. https://doi.org/10.1007/s11244-018-1070-7.

(110) Weatherup, R. S.; Wu, C. H.; Escudero, C.; Pérez-Dieste, V.; Salmeron, M. B. Environment-Dependent Radiation Damage in Atmospheric Pressure X-Ray Spectroscopy. *J Phys Chem B* **2018**, *122* (2), 737–744. https://doi.org/10.1021/acs.jpcb.7b06397.

(111) Nagy, Z.; You, H. Radiolytic Effects on the in Situ Investigation of Buried Interfaces with Synchrotron X-Ray Techniques. *Journal of Electroanalytical Chemistry* **1995**, *381* (1–2), 275–279. https://doi.org/10.1016/0022-0728(94)03772-U.

(112) Han, Y.; Axnanda, S.; Crumlin, E. J.; Chang, R.; Mao, B.; Hussain, Z.; Ross, P. N.; Li, Y.; Liu, Z. Observing the Electrochemical Oxidation of Co Metal at the Solid/Liquid Interface Using Ambient Pressure X-Ray Photoelectron Spectroscopy. *J Phys Chem B* **2018**, *122* (2), 666–671. https://doi.org/10.1021/acs.jpcb.7b05982.

(113) Favaro, M.; Yang, J.; Nappini, S.; Magnano, E.; Toma, F. M.; Crumlin, E. J.; Yano, J.; Sharp, I. D. Understanding the Oxygen Evolution Reaction Mechanism on $CoO_x$ Using Operando Ambient-Pressure X-Ray Photoelectron Spectroscopy. *J Am Chem Soc* **2017**, *139* (26), 8960–8970. https://doi.org/10.1021/jacs.7b03211.

(114) Ali-Löytty, H.; Louie, M. W.; Singh, M. R.; Li, L.; Sanchez Casalongue, H. G.; Ogasawara, H.; Crumlin, E. J.; Liu, Z.; Bell, A. T.; Nilsson, A.; Friebel, D. Ambient-Pressure XPS Study of a Ni–Fe Electrocatalyst for the Oxygen Evolution Reaction. *The Journal of Physical Chemistry C* **2016**, *120* (4), 2247–2253. https://doi.org/10.1021/acs.jpcc.5b10931.

(115) Favaro, M.; Drisdell, W. S.; Marcus, M. A.; Gregoire, J. M.; Crumlin, E. J.; Haber, J. A.; Yano, J. An Operando Investigation of (Ni–Fe–Co–Ce)$O_x$ System as Highly Efficient Electrocatalyst for Oxygen Evolution Reaction. *ACS Catal* **2017**, *7* (2), 1248–1258. https://doi.org/10.1021/acscatal.6b03126.

(116) Boucly, A.; Artiglia, L.; Fabbri, E.; Palagin, D.; Aegerter, D.; Pergolesi, D.; Novotny, Z.; Comini, N.; Diulus, J. T.; Huthwelker, T.; Ammann, M.; Schmidt, T. J. Direct Evidence of Cobalt Oxyhydroxide Formation on a $La_{0.2}Sr_{0.8}CoO_3$ Perovskite Water Splitting Catalyst. *J Mater Chem A Mater* **2022**, *10* (5), 2434–2444. https://doi.org/10.1039/D1TA04957G.

(117) Stoerzinger, K. A.; Favaro, M.; Ross, P. N.; Yano, J.; Liu, Z.; Hussain, Z.; Crumlin, E. J. Probing the Surface of Platinum during the Hydrogen Evolution Reaction in Alkaline Electrolyte. *J Phys Chem B* **2018**, *122* (2), 864–870. https://doi.org/10.1021/acs.jpcb.7b06953.

(118) Eilert, A.; Cavalca, F.; Roberts, F. S.; Osterwalder, J.; Liu, C.; Favaro, M.; Crumlin, E. J.; Ogasawara, H.; Friebel, D.; Pettersson, L. G. M.; Nilsson, A. Subsurface Oxygen in Oxide-Derived Copper Electrocatalysts for Carbon Dioxide Reduction. *J Phys Chem Lett* **2017**, *8* (1), 285–290. https://doi.org/10.1021/acs.jpclett.6b02273.

(119) Mom, R. v.; Falling, L. J.; Kasian, O.; Algara-Siller, G.; Teschner, D.; Crabtree, R. H.; Knop-Gericke, A.; Mayrhofer, K. J. J.; Velasco-Vélez, J.-J.; Jones, T. E. Operando Structure–Activity–Stability Relationship of Iridium Oxides during the Oxygen Evolution Reaction. *ACS Catal* **2022**, *12* (9), 5174–5184. https://doi.org/10.1021/acscatal.1c05951.

(120) Lichterman, M. F.; Hu, S.; Richter, M. H.; Crumlin, E. J.; Axnanda, S.; Favaro, M.; Drisdell, W.; Hussain, Z.; Mayer, T.; Brunschwig, B. S.; Lewis, N. S.; Liu, Z.; Lewerenz, H.-J. Direct Observation of the Energetics at a Semiconductor/Liquid Junction by Operando X-Ray Photoelectron Spectroscopy. *Energy Environ Sci* **2015**, *8* (8), 2409–2416. https://doi.org/10.1039/C5EE01014D.

(121) Shavorskiy, A.; Ye, X.; Karslıoğlu, O.; Poletayev, A. D.; Hartl, M.; Zegkinoglou, I.; Trotochaud, L.; Nemšák, S.; Schneider, C. M.; Crumlin, E. J.; Axnanda, S.; Liu, Z.; Ross, P. N.; Chueh, W.; Bluhm, H. Direct Mapping of Band Positions in Doped and Undoped Hematite during Photoelectrochemical Water Splitting. *J Phys Chem Lett* **2017**, *8* (22), 5579–5586. https://doi.org/10.1021/acs.jpclett.7b02548.

(122) Biesinger, M. C.; Payne, B. P.; Grosvenor, A. P.; Lau, L. W. M.; Gerson, A. R.; Smart, R. St. C. Resolving Surface Chemical States in XPS Analysis of First Row Transition Metals, Oxides and Hydroxides: Cr, Mn, Fe, Co and Ni. *Appl Surf Sci* **2011**, *257* (7), 2717–2730. https://doi.org/10.1016/j.apsusc.2010.10.051.

(123) van Elp, J.; Wieland, J. L.; Eskes, H.; Kuiper, P.; Sawatzky, G. A.; de Groot, F. M. F.; Turner, T. S. Electronic Structure of CoO, Li-Doped CoO, and $LiCoO_2$. *Phys Rev B* **1991**, *44* (12), 6090–6103. https://doi.org/10.1103/PhysRevB.44.6090.

(124) Reuter, K.; Scheffler, M. Surface Core-Level Shifts at an Oxygen-Rich Ru Surface: O/Ru(0001) vs. $RuO_2$(110). *Surf Sci* **2001**, *490* (1–2), 20–28. https://doi.org/10.1016/S0039-6028(01)01214-6.

(125) Zeng, Z.; Greeley, J. Characterization of Oxygenated Species at Water/Pt(111) Interfaces from DFT Energetics and XPS Simulations. *Nano Energy* **2016**, *29*, 369–377. https://doi.org/10.1016/j.nanoen.2016.05.044.

(126) Haverkort, M. W.; Zwierzycki, M.; Andersen, O. K. Multiplet Ligand-Field Theory Using Wannier Orbitals. *Phys Rev B* **2012**, *85* (16), 165113. https://doi.org/10.1103/PhysRevB.85.165113.

(127) Smekal, W.; Werner, W. S. M.; Powell, C. J. Simulation of Electron Spectra for Surface Analysis (SESSA): A Novel Software Tool for Quantitative Auger-Electron Spectroscopy and X-Ray Photoelectron Spectroscopy. *Surface and Interface Analysis* **2005**, *37* (11), 1059–1067. https://doi.org/10.1002/sia.2097.

(128) van der Veen, W. Checking "Hidden Layers" For Creating Greener Batteries And Catalysts. *University of Twente Newsroom*. 2021.

(129) Nemšák, S.; Strelcov, E.; Guo, H.; Hoskins, B. D.; Duchoň, T.; Mueller, D. N.; Yulaev, A.; Vlassiouk, I.; Tselev, A.; Schneider, C. M.; Kolmakov, A. In Aqua Electrochemistry Probed by XPEEM: Experimental Setup, Examples, and Challenges. *Top Catal* **2018**, *61* (20), 2195–2206. https://doi.org/10.1007/s11244-018-1065-4.

(130) Fabbri, E.; Nachtegaal, M.; Binninger, T.; Cheng, X.; Kim, B.-J.; Durst, J.; Bozza, F.; Graule, T.; Schäublin, R.; Wiles, L.; Pertoso, M.; Danilovic, N.; Ayers, K. E.; Schmidt, T. J. Dynamic Surface Self-Reconstruction Is the Key of Highly Active Perovskite Nano-Electrocatalysts for Water Splitting. *Nat Mater* **2017**, *16* (9), 925–931. https://doi.org/10.1038/nmat4938.

(131) Mefford, J. T.; Rong, X.; Abakumov, A. M.; Hardin, W. G.; Dai, S.; Kolpak, A. M.; Johnston, K. P.; Stevenson, K. J. Water Electrolysis on $La_{1-x}Sr_xCoO_{3-\delta}$ Perovskite Electrocatalysts. *Nat Commun* **2016**, *7*, 11053. https://doi.org/10.1038/ncomms11053.

(132) May, K. J.; Carlton, C. E.; Stoerzinger, K. A.; Risch, M.; Suntivich, J.; Lee, Y.-L.; Grimaud, A.; Shao-Horn, Y. Influence of Oxygen Evolution during Water Oxidation on the Surface of Perovskite Oxide Catalysts. *J Phys Chem Lett* **2012**, *3* (22), 3264–3270. https://doi.org/10.1021/jz301414z.

(133) Baeumer, C. Operando Characterization of Interfacial Charge Transfer Processes. *J Appl Phys* **2021**, *129* (17), 170901. https://doi.org/10.1063/5.0046142.

(134) Zhang, Y.; Khalifa, Y.; Maginn, E. J.; Newberg, J. T. Anion Enhancement at the Liquid–Vacuum Interface of an Ionic Liquid Mixture. *The Journal of Physical Chemistry C* **2018**, *122* (48), 27392–27401. https://doi.org/10.1021/acs.jpcc.8b07995.

(135) Pijolat, M.; Hollinger, G. New Depth-Profiling Method by Angular-Dependent x-Ray Photoelectron Spectroscopy. *Surf Sci* **1981**, *105* (1), 114–128. https://doi.org/10.1016/0039-6028(81)90151-5.

(136) Claessen, R.; Sing, M.; Paul, M.; Berner, G.; Wetscherek, A.; Müller, A.; Drube, W. Hard X-Ray Photoelectron Spectroscopy of Oxide Hybrid and Heterostructures: A New Method for the Study of Buried Interfaces. *New J Phys* **2009**, *11* (12), 125007. https://doi.org/10.1088/1367-2630/11/12/125007.

(137) Paynter, R. W. An ARXPS Primer. *J Electron Spectros Relat Phenomena* **2009**, *169* (1), 1–9. https://doi.org/10.1016/j.elspec.2008.09.005.

(138) Abruña, H. D.; Bommarito, G. M.; Acevedo, D. The Study of Solid/Liquid Interfaces with X-Ray Standing Waves. *Science (1979)* **1990**, *250* (4977), 69–74. https://doi.org/10.1126/science.250.4977.69.

(139) Martins, H. P.; Conti, G.; Cordova, I.; Falling, L.; Kersell, H.; Salmassi, F.; Gullikson, E.; Vishik, I.; Baeumer, C.; Naulleau, P.; Schneider, C. M.; Nemsak, S. Near Total Reflection X-Ray Photoelectron Spectroscopy: Quantifying Chemistry at Solid/Liquid and Solid/Solid Interfaces. *J Phys D Appl Phys* **2021**, *54* (46), 464002. https://doi.org/10.1088/1361-6463/ac2067.

(140) Ibach, H. *Physics of Surfaces and Interfaces*; Springer: Berlin, New York, 2006.

(141) Robinson, I. K. Crystal Truncation Rods and Surface Roughness. *Phys Rev B* **1986**, *33* (6), 3830–3836. https://doi.org/10.1103/PhysRevB.33.3830.

(142) Barbier, A.; Mocuta, C.; Kuhlenbeck, H.; Peters, K. F.; Richter, B.; Renaud, G. *Atomic Structure of the Polar NiO(111)-p 2 3 2 Surface*; 2000.

(143) Vonk, V.; Konings, S.; van Hummel, G. J.; Harkema, S.; Graafsma, H. The Atomic Surface Structure of SrTiO3(0 0 1) in Air Studied with Synchrotron X-Rays. *Surf Sci* **2005**, *595* (1–3), 183–193. https://doi.org/10.1016/j.susc.2005.08.010.

(144) Feidenhansl, R.; Grey, F.; Johnson, R. L.; Mochrie, S. G. J.; Bohr, J.; Nielsen, M. Oxygen Chemisorption on Cu(110): A Structural Determination by x-Ray Diffraction. *Phys Rev B* **1990**, *41* (8). https://doi.org/10.1103/PhysRevB.41.5420.

(145) Yacoby, Y.; Zhou, H.; Pindak, R.; Božović, I. Atomic-Layer Synthesis and Imaging Uncover Broken Inversion Symmetry in La2-XSrxCuO4 Films. *Phys Rev B Condens Matter Mater Phys* **2013**, *87* (1). https://doi.org/10.1103/PhysRevB.87.014108.

(146) Willmott, P. R.; Pauli, S. A.; Herger, R.; Schlepütz, C. M.; Martoccia, D.; Patterson, B. D.; Delley, B.; Clarke, R.; Kumah, D.; Cionca, C.; Yacoby, Y. Structural Basis for the Conducting Interface between $LaAlO_3$ and $SrTiO_3$. *Phys Rev Lett* **2007**, *99* (15).

(147) Liu, Y.; Barbour, A.; Komanicky, V.; You, H. X-Ray Crystal Truncation Rod Studies of Surface Oxidation and Reduction on Pt(111). *Journal of Physical Chemistry C* **2016**, *120* (29), 16174–16178. https://doi.org/10.1021/acs.jpcc.6b00492.

(148) Eng, P. J.; Trainor, T. P.; Brown, G. E.; Waychunas, G. A.; Newville, M.; Sutton, S. R.; Rivers, M. L. Structure of the Hydrated α-Al2O3 (0001) Surface. *Science (1979)* **2000**, *288* (5468). https://doi.org/10.1126/science.288.5468.1029.

(149) Magnussen, O. M.; Krug, K.; Ayyad, A. H.; Stettner, J. In Situ Diffraction Studies of Electrode Surface Structure during Gold Electrodeposition. *Electrochim Acta* **2008**, *53* (9). https://doi.org/10.1016/j.electacta.2007.10.037.

(150) Harlow, G. S.; Aldous, I. M.; Thompson, P.; Gründer, Y.; Hardwick, L. J.; Lucas, C. A. Adsorption, Surface Relaxation and Electrolyte Structure at Pt(111) Electrodes in Non-Aqueous and Aqueous Acetonitrile Electrolytes. *Physical Chemistry Chemical Physics* **2019**, *21* (17), 8654–8662. https://doi.org/10.1039/C9CP00499H.

(151) Baggio, B. F.; Grunder, Y. In Situ X-Ray Techniques for Electrochemical Interfaces. *Annual Review of Analytical Chemistry*. 2021. https://doi.org/10.1146/annurev-anchem-091020-100631.

(152) Harlow, G. S.; Lundgren, E.; Escudero-Escribano, M. Recent Advances in Surface X-Ray Diffraction and the Potential for Determining Structure-Sensitivity Relations in Single-Crystal Electrocatalysis. *Curr Opin Electrochem* **2020**, *23*, 162–173. https://doi.org/10.1016/j.coelec.2020.08.005.

(153) Gründer, Y.; Lucas, C. A. Surface X-Ray Diffraction Studies of Single Crystal Electrocatalysts. *Nano Energy* **2016**, *29*, 378–393. https://doi.org/10.1016/j.nanoen.2016.05.043.

(154) Escudero-Escribano, M.; Pedersen, A. F.; Ulrikkeholm, E. T.; Jensen, K. D.; Hansen, M. H.; Rossmeisl, J.; Stephens, I. E. L.; Chorkendorff, I. Active-Phase Formation and Stability of Gd/Pt(111) Electrocatalysts for Oxygen Reduction: An In Situ Grazing Incidence X-Ray Diffraction Study. *Chemistry - A European Journal* **2018**, *24* (47). https://doi.org/10.1002/chem.201801587.

(155) You, H.; Zurawski, D. J.; Nagy, Z.; Yonco, R. M. In-Situ x-Ray Reflectivity Study of Incipient Oxidation of Pt(111) Surface in Electrolyte Solutions. *J Chem Phys* **1994**, *100* (6). https://doi.org/10.1063/1.466254.

(156) Tidswell, I. M.; Markovic, N. M.; Ross, P. N. Potential Dependent Surface Structure of the Pt(1 1 1) Electrolyte Interface. *Journal of Electroanalytical Chemistry* **1994**, *376* (1–2). https://doi.org/10.1016/0022-0728(94)03553-9.

(157) Ruge, M.; Drnec, J.; Rahn, B.; Reikowski, F.; Harrington, D. A.; Carlà, F.; Felici, R.; Stettner, J.; Magnussen, O. M. Electrochemical Oxidation of Smooth and Nanoscale Rough Pt(111): An In Situ Surface X-Ray Scattering Study. *J Electrochem Soc* **2017**, *164* (9). https://doi.org/10.1149/2.0741709jes.

(158) Drnec, J.; Harrington, D. A.; Magnussen, O. M. Electrooxidation of Pt(111) in Acid Solution. *Current Opinion in Electrochemistry*. 2017. https://doi.org/10.1016/j.coelec.2017.09.021.

(159) Fuchs, T.; Drnec, J.; Calle-Vallejo, F.; Stubb, N.; Sandbeck, D. J. S.; Ruge, M.; Cherevko, S.; Harrington, D. A.; Magnussen, O. M. Structure Dependency of the Atomic-Scale Mechanisms of Platinum Electro-Oxidation and Dissolution. *Nat Catal* **2020**, *3* (9), 754–761. https://doi.org/10.1038/s41929-020-0497-y.

(160) Weber, T.; Vonk, V.; Escalera-López, D.; Abbondanza, G.; Larsson, A.; Koller, V.; Abb, M. J. S.; Hegedüs, Z.; Bäcker, T.; Lienert, U.; Harlow, G. S.; Stierle, A.; Cherevko, S.; Lundgren, E.; Over, H. Operando Stability Studies of Ultrathin Single-Crystalline IrO2(110) Films under Acidic Oxygen Evolution Reaction Conditions. *ACS Catal* **2021**, *11* (20). https://doi.org/10.1021/acscatal.1c03599.

(161) Chu, Y. S.; Lister, T. E.; Cullen, W. G.; You, H.; Nagy, Z. Commensurate Water Monolayer at the RuO2 (110)/Water Interface. *Phys Rev Lett* **2001**, *86* (15), 3364–3367. https://doi.org/10.1103/PhysRevLett.86.3364.

(162) Brown, G. E.; Parks, G. A. Sorption of Trace Elements on Mineral Surfaces: Modern Perspectives from Spectroscopic Studies, and Comments on Sorption in the Marine Environment. *Int Geol Rev* **2001**, *43* (11). https://doi.org/10.1080/00206810109465060.

(163) Stumm, W. Chemistry of the Solid-Water Interface: Processes at the Mineral- Water and Particle-Water Interface in Natural Systems. *Chemistry of the solid-water interface: processes at the mineral- water and particle-water interface in natural systems* **1992**. https://doi.org/10.1097/00010694-199309000-00010.

(164) Raj, H. P. C. R. *Principles of Colloid and Surface Chemistry, Revised and Expanded*; 1997.

(165) Brown, G. E.; Sturchio, N. C. An Overview of Synchrotron Radiation Applications to Low Temperature Geochemistry and Environmental Science. *Reviews in Mineralogy and Geochemistry*. 2002. https://doi.org/10.2138/gsrmg.49.1.1.

(166) Garcia, A. C.; Touzalin, T.; Nieuwland, C.; Perini, N.; Koper, M. T. M. Enhancement of Oxygen Evolution Activity of Nickel Oxyhydroxide by Electrolyte Alkali Cations. *Angewandte Chemie - International Edition* **2019**, *58* (37), 12999–13003. https://doi.org/10.1002/anie.201905501.

(167) Yang, C.; Fontaine, O.; Tarascon, J.-M.; Grimaud, A. Chemical Recognition of Active Oxygen Species on the Surface of Oxygen Evolution Reaction Electrocatalysts. *Angewandte Chemie* **2017**, *129* (30). https://doi.org/10.1002/ange.201701984.

(168) Strmcnik, D.; Kodama, K.; van der Vliet, D.; Greeley, J.; Stamenkovic, V. R.; Marković, N. M. The Role of Non-Covalent Interactions in Electrocatalytic Fuel-Cell Reactions on Platinum. *Nat Chem* **2009**, *1* (6). https://doi.org/10.1038/nchem.330.

(169) Ledezma-Yanez, I.; Wallace, W. D. Z.; Sebastián-Pascual, P.; Climent, V.; Feliu, J. M.; Koper, M. T. M. Interfacial Water Reorganization as a PH-Dependent Descriptor of the Hydrogen Evolution Rate on Platinum Electrodes. *Nat Energy* **2017**, *2* (4). https://doi.org/10.1038/nenergy.2017.31.

(170) Rossmeisl, J.; Chan, K.; Skúlason, E.; Björketun, M. E.; Tripkovic, V. On the PH Dependence of Electrochemical Proton Transfer Barriers. *Catal Today* **2016**, *262*. https://doi.org/10.1016/j.cattod.2015.08.016.

(171) MacHesky, M.; Wesolowski, D.; Rosenqvist, J.; Předota, M.; Vlcek, L.; Ridley, M.; Kohli, V.; Zhang, Z.; Fenter, P.; Cummings, P.; Lvov, S.; Fedkin, M.; Rodriguez-Santiago, V.; Kubicki, J.; Bandura, A. Comparison of Cation Adsorption by Isostructural Rutile and Cassiterite. *Langmuir* **2011**, *27* (8). https://doi.org/10.1021/la1040163.

(172) Předota, M.; Zhang, Z.; Fenter, P.; Wesolowski, D. J.; Cummings, P. T. Electric Double Layer at the Rutile (110) Surface. 2. Adsorption of Ions from Molecular Dynamics and X-Ray Experiments. *Journal of Physical Chemistry B* **2004**, *108* (32). https://doi.org/10.1021/jp037199x.

(173) Kawaguchi, T.; Liu, Y.; Reiter, A.; Cammarot, C.; Pierce, M. S.; You, H. Direct Determination of One-Dimensional Interphase Structures Using Normalized Crystal Truncation Rod Analysis. *J Appl Crystallogr* **2018**, *51*. https://doi.org/10.1107/S1600576718004326.

(174) Liu, Y.; Kawaguchi, T.; Pierce, M. S.; Komanicky, V.; You, H. Layering and Ordering in Electrochemical Double Layers. *Journal of Physical Chemistry Letters* **2018**, *9* (6). https://doi.org/10.1021/acs.jpclett.8b00123.

(175) Kawaguchi, T.; Rao, R. R.; Lunger, J. R.; Liu, Y.; Walko, D.; Karapetrova, E. A.; Komanicky, V.; Shao-Horn, Y.; You, H. Stern Layers on RuO2 (100) and (110) in Electrolyte: Surface X-Ray Scattering Studies. *Journal of Electroanalytical Chemistry* **2020**, *875*. https://doi.org/10.1016/j.jelechem.2020.114228.

(176) Rao, R. R.; Huang, B.; Katayama, Y.; Hwang, J.; Kawaguchi, T.; Lunger, J. R.; Peng, J.; Zhang, Y.; Morinaga, A.; Zhou, H.; You, H.; Shao-Horn, Y. PH-and Cation-Dependent Water Oxidation on Rutile RuO2(110). *Journal of Physical Chemistry C* **2021**, *125* (15). https://doi.org/10.1021/acs.jpcc.1c00413.

(177) Hussain, H.; Tocci, G.; Woolcot, T.; Torrelles, X.; Pang, C. L.; Humphrey, D. S.; Yim, C. M.; Grinter, D. C.; Cabailh, G.; Bikondoa, O.; Lindsay, R.; Zegenhagen, J.; Michaelides, A.; Thornton, G. Structure of a Model TiO2 Photocatalytic Interface. *Nat Mater* **2017**, *16* (4). https://doi.org/10.1038/NMAT4793.

(178) Vlieg, E. ROD: A Program for Surface X-Ray Crystallography. *Journal of Applied Crystallography*. 2000. https://doi.org/10.1107/S0021889899013655.

(179) Björck, M.; Andersson, G. GenX: An Extensible X-Ray Reflectivity Refinement Program Utilizing Differential Evolution. *J Appl Crystallogr* **2007**, *40* (6), 1174–1178. https://doi.org/10.1107/S0021889807045086.

(180) Joly, Y.; Abisset, A.; Bailly, A.; de Santis, M.; Fettar, F.; Grenier, S.; Mannix, D.; Ramos, A. Y.; Saint-Lager, M. C.; Soldo-Olivier, Y.; Tonnerre, J. M.; Guda, S. A.; Gründer, Y. Simulation of Surface Resonant X-Ray Diffraction. *J Chem Theory Comput* **2018**, *14* (2). https://doi.org/10.1021/acs.jctc.7b01032.

(181) Laanait, N.; Zhang, Z.; Schlepütz, C. M. Imaging Nanoscale Lattice Variations by Machine Learning of X-Ray Diffraction Microscopy Data. *Nanotechnology* **2016**, *27* (37). https://doi.org/10.1088/0957-4484/27/37/374002.

(182) Ning, Y.; Li, Y.; Wang, C.; Li, R.; Zhang, F.; Zhang, S.; Wang, Z.; Yang, F.; Zong, N.; Peng, Q.; Xu, Z.; Wang, X.; Li, R.; Breitschaft, M.; Hagen, S.; Schaff, O.; Fu, Q.; Bao, X. Tunable Deep Ultraviolet Laser Based near Ambient Pressure Photoemission Electron Microscope for Surface Imaging in the Millibar Regime. *Review of Scientific Instruments* **2020**, *91* (11), 113704. https://doi.org/10.1063/5.0016242.

(183) Roy, K.; Raabe, J.; Schifferle, P.; Finizio, S.; Kleibert, A.; van Bokhoven, J. A.; Artiglia, L. Design and Performance of a New Setup for Spatially Resolved Transmission X-Ray Photoelectron Spectroscopy at the Swiss Light Source. *J Synchrotron Radiat* **2019**, *26* (3), 785–792. https://doi.org/10.1107/S1600577519002984.

(184) Roth, F.; Neppl, S.; Shavorskiy, A.; Arion, T.; Mahl, J.; Seo, H. O.; Bluhm, H.; Hussain, Z.; Gessner, O.; Eberhardt, W. Efficient Charge Generation from Triplet Excitons in Metal-Organic Heterojunctions. *Phys Rev B* **2019**, *99* (2), 020303. https://doi.org/10.1103/PhysRevB.99.020303.

(185) Brüninghoff, R.; Wenderich, K.; Korterik, J. P.; Mei, B. T.; Mul, G.; Huijser, A. Time-Dependent Photoluminescence of Nanostructured Anatase TiO 2 and the Role of Bulk and Surface Processes. *The Journal of Physical Chemistry C* **2019**, *123* (43), 26653–26661. https://doi.org/10.1021/acs.jpcc.9b06890.

(186) Huijser, A.; Pan, Q.; van Duinen, D.; Laursen, M. G.; el Nahhas, A.; Chabera, P.; Freitag, L.; González, L.; Kong, Q.; Zhang, X.; Haldrup, K.; Browne, W. R.; Smolentsev, G.; Uhlig, J. Shedding Light on the Nature of Photoinduced States Formed in a Hydrogen-Generating Supramolecular RuPt Photocatalyst by Ultrafast Spectroscopy. *J Phys Chem A* **2018**, *122* (31), 6396–6406. https://doi.org/10.1021/acs.jpca.8b00916.

(187) Kumar, S.; Graves, C. E.; Strachan, J. P.; Kilcoyne, A. L. D.; Tyliszczak, T.; Nishi, Y.; Williams, R. S. In-Operando Synchronous Time-Multiplexed O K-Edge x-Ray Absorption Spectromicroscopy of Functioning Tantalum Oxide Memristors. *J Appl Phys* **2015**, *118* (3), 034502. https://doi.org/10.1063/1.4926477.

(188) Kersell, H.; Chen, P.; Martins, H.; Lu, Q.; Brausse, F.; Liu, B.-H.; Blum, M.; Roy, S.; Rude, B.; Kilcoyne, A.; Bluhm, H.; Nemšák, S. Simultaneous Ambient Pressure X-Ray Photoelectron Spectroscopy and Grazing Incidence X-Ray Scattering in Gas Environments. **2021**, 1–20.